\documentstyle[12pt,aasms4]{article}
\input{psfig.tex}  

\begin{document}

\title{ AGES OF GLOBULAR CLUSTERS FROM HIPPARCOS PARALLAXES OF LOCAL
SUBDWARFS
\footnote{Based on data from the Hipparcos astrometry satellite and from Asiago
and McDonald Observatories} }

\author{Raffaele G. Gratton}
\affil{Osservatorio Astronomico di Padova, Vicolo dell'Osservatorio 5, 35122
  Padova, ITALY\\gratton@pdmida.pd.astro.it}
\authoremail{gratton@pdmida.pd.astro.it}

\author{Flavio Fusi Pecci}
\affil{Osservatorio Astronomico di Bologna, ITALY, and
Stazione Astronomica, 09012 Capoterra, Cagliari, ITALY\\
flavio@astbo3.bo.astro.it}
\authoremail{flavio@astbo3.bo.astro.it}

\author{Eugenio Carretta and Gisella Clementini}
\affil{Osservatorio Astronomico di Bologna, ITALY\\
carretta@astbo3.bo.astro.it and gisella@astbo3.bo.astro.it}
\authoremail{carretta@astbo3.bo.astro.it}
\authoremail{gisella@astbo3.bo.astro.it}

\author{Carlo E. Corsi}
\affil{Osservatorio Astronomico di Roma, ITALY\\
corsi@oarhp4.rm.astro.it}
\authoremail{corsi@oarhp4.rm.astro.it}

\author{Mario Lattanzi}
\affil{Osservatorio Astronomico di Torino, ITALY\\
lattanzi@gsc2.to.astro.it}
\authoremail{lattanzi@gsc2.to.astro.it}

\begin{abstract}

We report here initial but still strongly conclusive results on {\it absolute}
ages of galactic globular clusters (GGCs). This study is based on high
precision trigonometric parallaxes from the HIPPARCOS satellite coupled with
accurate metal abundances ([Fe/H], [O/Fe], and [$\alpha$/Fe]) from high
resolution spectroscopy for a sample of about thirty subdwarfs. Systematic
effects due to star selection (Lutz-Kelker corrections to parallaxes) and to
the possible presence of undetected binaries in the sample of {\it bona fide}
single stars are examined, and appropriate corrections are estimated. They are
found to be small for our sample. The new data allowed us to reliably define
the {\it absolute} location of the main sequence (MS) as a function of
metallicity. 

These results are then used to derive distances and ages for a carefully
selected sample of nine globular clusters having metallicities determined from
high dispersion spectra of individual giants according to a procedure totally
consistent with that used for the field subdwarfs. Very precise and homogeneous
reddening values have also been independently determined for these clusters.
Random errors of our distance moduli are $\pm 0.08$~mag, and systematic errors
are likely of the same order of magnitude. These very accurate distances allow
us to derive ages with internal errors of $\sim 12\%$\ ($\pm 1.5$~Gyr). 

The main results are: 
\begin{itemize}

\item HIPPARCOS parallaxes are smaller than the corresponding ground-based 
measurements leading, in turn, to longer distance moduli ($\sim 0.2$ mag)
and younger ages ($\sim 2.8$ Gyr).

\item The distance to NGC6752 derived from our MS-fitting is consistent with
that determined using the white dwarf cooling sequence. 

\item The relation between the zero age HB (ZAHB) absolute magnitude and
metallicity for the nine programme clusters turns out to be: 
$$ M_V(ZAHB) = (0.22\pm 0.09)({\rm [Fe/H]}+1.5) + (0.49\pm 0.04)$$
This relation is fairly consistent with some of the most recent theoretical
models. Within quoted errors, the slope is in agreement with that given by the
Baade-Wesselink (BW) analysis of RR Lyraes (Fernley 1994, Clementini et al.
1995), while it is somewhat shallower than the relation given by Sandage
(1993). The zero point is 0.2 to 0.3~mag brighter than that obtained with BW,
while it agrees fairly well with that given by Sandage. Comparison with
alternative relationships is briefly discussed. 

\item The corresponding LMC distance modulus is $(m-M)_0=18.60\pm 0.07$,
in good agreement with the recent values of $18.70\pm 0.10$ and $18.54\pm 0.2$\
derived by Feast and Catchpole (1997) and van Leeuwen et al. (1997),
respectively, from HIPPARCOS parallaxes of Galactic Cepheid and Mira 
variables.

\item The age of the {\it bona fide} old globular clusters (Oosterhoff II and 
BHB) based on the absolute magnitude of the turn-off, a theoretically 
robust indicator, is:
$${\rm Age} = {11.8^{+2.1}_{-2.5}} {\rm Gyr}$$
(where the error bar is the 95\% confidence range). The r.m.s. scatter of
individual ages around the mean value is $\sim 10\%$, in agreement with
expectations from observational errors alone (that is, we do not find necessary
to introduce a real age scatter amongst these clusters). A reliable study of
the {\it relative} ages requires the use of age indicators better suited for
this purpose and data for a larger sample of GGCs. 

\item Allowing for a minimum delay of 0.5 Gyr from the birth of the Universe
before the formation of globular clusters, our age estimate is compatible with
an Einstein-de Sitter model if $H_0\leq 64$ km s$^{-1}$Mpc$^{-1}$, or $H_0\leq
83$ km s$^{-1}$Mpc$^{-1}$\ in a flat universe with $\Omega_m=0.2$. Since these
upper limits are well within the confidence range of most determinations of
$H_0$, we conclude that the present age of globular clusters does not conflict
with standard inflationary models for the Universe. 
\end{itemize}

\end{abstract}

\keywords{ Clusters: globulars -- Cosmology -- Stars: basic parameters --
Stars: stellar models } 

\newpage

\section{INTRODUCTION}

In the density perturbations that later became galaxies, globular clusters
(GCs) were almost the first objects to be formed, only preceeded by the very
first generation of near zero-metallicity stars. GCs are indeed the
oldest objects in our own Galaxy that can be dated with some precision,
exploiting the expectations of the stellar evolution theory. Their ages
provide then a stringent lower limit to the age $t$\ of the Universe. 

In a flat universe (and for the range of parameters most likely to be correct), 
$t$\ is roughly related to the Hubble constant $H_0$\ and the energy density 
parameter $\Omega_m$ by the relation: 
\begin{equation}
\Omega_m-0.7 \Omega_\Lambda \simeq 5.8 (1-1.3 h t),  
\end{equation}
where $\Omega_\Lambda=\Lambda c^2/3 H^2$, $\Lambda$\ being the cosmological
constant, $h=H_0/100$~km s$^{-1}$, and $t$ is in units of $10^{10}$~yr (see
e.g. Dekel, Burstein \& White 1996). The simplest model is the Einstein-de
Sitter model ($\Omega_m=1$\ and $\Omega_\Lambda=0$). A generalization of this
relation still compatible with standard inflationary model is a flat universe
with $\Omega_m+\Omega_\Lambda=1$, where $\Omega_m$\ can be smaller than unity
but only at the expense of a non zero cosmological constant. 

The status of knowledge of measures of $\Omega$\ is summarized by Dekel et al.
(1996): the estimates based on virialized objects typically yield low values of
$\Omega_m\sim 0.2-0.3$, while global measures typically indicate higher values
($\Omega_m\sim 0.4-1$). However, for various reasons it is possible that the
first group of measures is underestimated. Anyway, we will assume a lower
limit of $\Omega_m>0.2$. 

As to the Hubble constant, the debate on its value is still open (see the
review by Kennicutt, Freedman, \& Mould 1995). However, most recent
determinations are in the range $H_0\sim 55-75$  km s$^{-1}$Mpc$^{-1}$\ (see
e.g. Freedman et al. 1997, $H_0=73\pm 10$ km s$^{-1}$Mpc$^{-1}$\ from Cepheids;
Hamuy et al. 1996, $H_0=63.1\pm 3.4\pm 2.9$ km s$^{-1}$Mpc$^{-1}$\ from type Ia
SNe; Saha et al. 1997, $H_0=58\pm 7 $ km s$^{-1}$Mpc$^{-1}$\ from HST-SNe;
Sandage \& Tammann 1997, $H_0=56\pm 7 $ km s$^{-1}$Mpc$^{-1}$). With this value
for $H_0$, the age of the universe is constrained to be $t<11.6$~Gyr in an
Einstein-de Sitter model, and $t<14.9$~Gyr in a flat universe with
$\Omega_m=0.2$. 
 
Recent determinations for the age of globular clusters are those by Bolte \&
Hogan (1995), who found t=$15.8\pm 2.1$~Gyr; by Chaboyer (1995), who found an
age range from 11 to 21~Gyr (more recently, Chaboyer et al. 1996 suggested a
more restricted range: t=$14.6\pm 2.5$~Gyr); and by VandenBerg, Bolte \&
Stetson (1996), t=$15^{+5}_{-3}$~Gyr. If allowance is given for at least
0.5~Gyr for the globular cluster formation after the Big Bang, the age found by
Bolte and Hogan is not compatible with an Einstein-de Sitter universe unless
$H_0 \sim 40$ km s$^{-1}$Mpc$^{-1}$\ (out of the confidence range of most
determinations) or with flat models with $\Omega_m=0.2$ unless $H_0 \sim 58$ km
s$^{-1}$Mpc$^{-1}$. The somewhat lower value of VandenBerg, Bolte \& Stetson
(1996) still requires $H_0<50$ km s$^{-1}$Mpc$^{-1}$\ for an Einstein-de Sitter
universe, and $H_0<65$ km s$^{-1}$Mpc$^{-1}$\ in a flat universe with
$\Omega_m=0.2$. Finally, even the larger error bar of Chaboyer (1995) can
accommodate an Einstein-de Sitter model only if $H_0<55$~km s$^{-1}$Mpc$^{-1}$.
This discrepancy could be reconciled if globular cluster ages based on the
recent horizontal branch (HB) models by Mazzitelli, D'Antona \& Caloi 1995;
Salaris, Degl'Innocenti \& Weiss 1997; and D'Antona, Caloi \& Mazzitelli 1997,
were considered. However, other recent models (VandenBerg 1997) give fainter
HBs, so that there is no general consensus about these age determinations. 

Given the large errors in the Hubble constant, some range of overlap
can still be found between globular cluster ages and standard inflationary
models for the Universe; however it is difficult to escape the impression that
this overlap is very small, and that current ages for globular clusters are
indeed uncomfortably large in the framework of standard cosmological models. 

The main difficulty in the derivation of the ages of globular clusters is the
large uncertainty in the distance scale (Renzini 1991). Distances toward
globular clusters can be derived using various techniques. Most determinations
depend on stellar models, like e.g. main sequence fitting to isochrones or
those based on the comparison with horizontal branch (HB) models mentioned
above. However, distances determined by these procedures are hampered by the
uncertainties still existing in the equation of state, in the treatment of
convection, and in the transformations from the theoretical
$\log{L/L_\odot}-\log{T_{\rm eff}}$\ plane to the observational $M_V-color$\
plane (leaving aside issues like core rotation, He-diffusion, or WIMPS: see
VandenBerg et al. 1996). On the other side, in principle there are at least
three different feasible techniques that can provide distances to globular
clusters with the accuracy ($\sim 0.1$~mag) required for accurate dating and do
not depend too heavily on stellar models: 

\begin{itemize}

\item The Baade-Wesselink (BW) method may provide direct estimates of the 
absolute magnitude of cluster pulsating variables. So far, this technique 
has been applied to about 30 field RR Lyraes and to a few variables in
three GGCs, namely M92 and M5 (Storm, Carney \& Latham 1994) and M4 
(Liu and Janes 1990). However, while the BW method seems to give a reliable 
ranking of distances (Carney, Storm \& Jones 1992), there are serious 
suspicions about its zero-point (Walker 1992, Fernley 1994, Carney et al.
1995). Indeed, the distance to the LMC obtained from the absolute
magnitudes of RR Lyrae variables calibrated against the results of the 
BW method disagrees with the distance derived using classical Cepheids, the 
circumstellar ring of SN1987A, and the Mira PL relation
(Walker 1992). The discrepancy is even more striking if one uses the latest
results from the HIPPARCOS satellite (Feast and Catchpole 1997). 

\item The cooling sequence of white dwarfs provides a faint but theoretically
rather secure distance ladder, independent of metallicity and details of the
convection theory (but dependent on the observational assumptions related to
colors). This technique (Fusi Pecci and Renzini 1979) has been recently 
applied to NGC 6752 using HST data (Renzini et al. 1996). The procedure 
assumes that the observed white dwarfs are DA, but direct verification of 
this assumption is beyond present spectroscopic capabilities. Furthermore,
the white dwarf distance estimates depend rather critically on the
assumption that the calibrating white dwarfs have the same mass of the GC
white dwarfs, an assumption that may be criticized in view of the differences
in the age of the parent populations.

\item Traditionally, the simplest technique to derive distances to clusters is
to compare their main sequence (MS) with a suitable template (Sandage 1970).
Unfortunately, while this procedure works very well with (population I) open
clusters, the template main sequence for metal-poor globular clusters has been
up to now quite uncertain due to the paucity of metal-poor dwarfs in the solar
neighborhood. Hence, very few reliable subdwarf parallaxes could be measured
from the ground, and systematic errors are likely to exist given the low S/N
ratio obtained and the relevance of the observational biases. 

\end{itemize}

The measure of accurate distances to a large number of subdwarfs, to be used in
the determination of distances to globular clusters, has been one of the major
targets of the HIPPARCOS mission (Perryman et al. 1989). In this paper, we
present the results of distance and age derivations for the nine best observed
globular clusters, based on the HIPPARCOS parallaxes for a sample of nearby
dwarfs with $-2.5<$[Fe/H]$<-0.5$. For brevity, only  the main features of our
analysis will be presented here, while details of the method and tabular
presentations of the whole database will be given in forthcoming papers. 

\section{BASIC DATA FOR SUBDWARFS}

\subsection{Parallaxes}

HIPPARCOS has provided absolute parallaxes for over 118,000 stars, with typical
accuracies (standard errors) of $\sim 1$ mas (to be compared with best case
errors of 2-3 mas obtained through painstaking efforts from the ground).
Several hundred metal-poor dwarfs are in the HIPPARCOS input catalogue. Within
the FAST proposal n. 022, we had early access to data for a sample of 99 
subdwarfs. The present final version of this paper also includes data for
additional stars available after the Hipparcos database was realeased to
the astronomical community; part of these stars are from the list by Reid
(1997). Reid's list includes more metal-poor stars than those considered
here, however most of these stars are quite distant, and hence have large
relative errors in the parallaxes. Using his data we may add 5 stars to our
list, and improve the parallax for a sixth one (HD188510). Three of these
additional stars are reddened, according to Carney et al. (1994); this will be
considered in the following discussion. A few other stars having high precision
abundances were also considered. 

In Table~1, we list the HIPPARCOS parallaxes and colors for the subset of stars
in our sample that have $\Delta\pi/\pi<0.12$\ (where $\Delta\pi$\ is the
standard error) and $5<M_V<8$; these are the stars most useful for the present
purposes (see discussion below). Data for both single and binary stars are
given in Table~1: however, only bona fide single stars with $M_V>5.5$ will be
used in the derivation of GC ages. Note that errors in the derived absolute
magnitudes are $\leq 0.25$~mag. 

\begin{figure} 
\centerline{\hbox{\psfig{figure=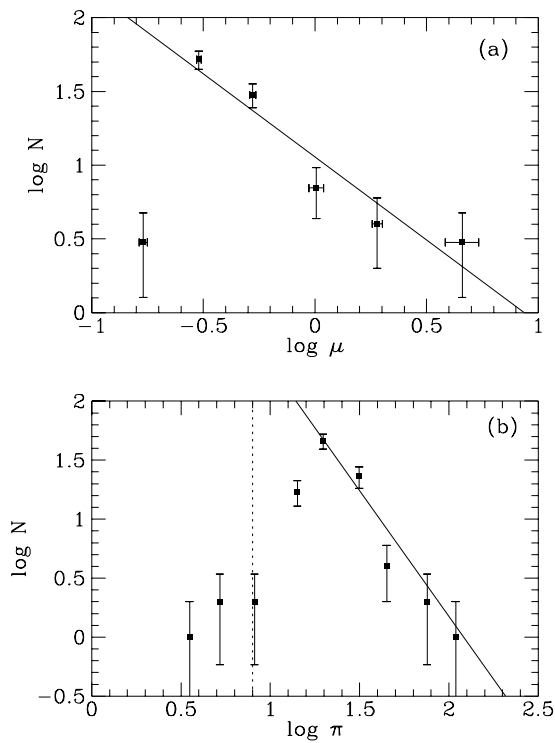,width=13.0cm,clip=}}}  
\medskip                                                                       
\caption{ (a) Distribution of proper motions for the stars in our original
HIPPARCOS sample. The best fit line with slope $-1.13$\ is shown overimposed.
(b) Distribution of parallaxes for the stars in our original HIPPARCOS sample.
Solid line is the power law with slope $-2.13$\ expected from the proper motion
distribution. The dashed line represents the approximate value of the adopted
threshold. Note that the threshold is well below the lower extreme of the
region of validity of the power law distribution } 
\label{f:fig1} 
\end{figure}

The parallaxes and the absolute magnitudes $M_V$\ for the stars in our original
sample (first part of Table~1) do not include any Lutz-Kelker correction (Lutz
\& Kelker 1973). This is a statistical correction (i.e. it applies only to the
average values: the parallaxes listed in the HIPPARCOS catalogue are in any
case the best estimate for the parallax of individual stars), and should take
into account two systematic effects: (i) stars with parallaxes measured too
high have a higher probability than those with parallaxes measured too low to
be included in the sample; and (ii) more weight (roughly proportional to
$(\pi/\Delta\pi)^2$) is given to stars with parallaxes that are overestimated
rather than to stars with underestimated parallaxes. The value of the
corrections to be applied depends on the distribution of the true parallaxes
$\pi_t$\ in the original sample (in our case, the list of 99 stars selected
before HIPPARCOS parallaxes were known) and on the observational errors. In
principle the true parallax distribution could be derived from the observed
distribution, after deconvolution for the observational error. However, the
deconvolution process is quite uncertain; hence most authors (see e.g.
VandenBerg et al. 1996, Sandquist et al. 1996, and Reid 1997) prefer to use the
procedure by Hanson (1979) \footnote{Note that standard errors for parallaxes
should be used in Hanson formulae; these are the errors quoted by HIPPARCOS and
by the 1991 version of the General Catalogue of Trigonometric Parallaxes (van
Altena, Lee \& Hoffleit 1991).}, who gave analytical formulae relating the
absolute magnitude corrections to the distribution of proper motions $\mu$,
which is modeled as a power law with exponent $x$. $x$\ is related to the
exponent $n$\ of the analogous power law for the parallax distribution by the
simple relation $n=x-1$. If we consider only stars with $\mu>0.2$~arcsec/yr
(only 4 out of 99 stars in our sample have $\mu<0.2$~arcsec/yr), proper motions
for our sample distribute as $N(\mu)\sim \mu^{-1.13\pm 0.20}$\ (see panel {\it
a} of Figure~\ref{f:fig1}). The appropriate Lutz-Kelker corrections to be used
would then be: 
\begin{equation}
\Delta M_{LK}=-5.71 (\Delta \pi/\pi)^2-27.95 (\Delta \pi/\pi)^4
\end{equation}
Weighted average corrections to the absolute magnitude obtained with this
formula are about $\Delta M_{LK}=-0.012$~mag (corresponding to ages younger by
$\sim 0.18$~Gyr), with individual values in the range $-0.06\leq \Delta
M_{LK}\leq 0$. However, as noted by Hanson (1979), this procedure is correct
only if the whole parallax distribution (both above and below the threshold) is
represented by a single power law. This is not the case for our sample: in fact
we found that the above mentioned power law is appropriate only for parallaxes
larger than $\sim 15$~mas, well above the threshold ($\sim 8$~mas: see panel
{\it b} of Figure~\ref{f:fig1}). In fact, 89 out of 99 stars in our sample have
a measured parallax above the threshold. This means that near the threshold,
the exponent $n$\ of the parallax distribution is strongly positive;
unfortunately too few stars are available for a reliable determination. Anyway,
Eq. (2) is clearly inappropriate, since it overestimates the Lutz-Kelker
corrections most appropriate for our sample. We obtain a more reliable estimate
by means of MonteCarlo simulations. Our procedure was as follows: first, we
assumed that the observed parallax distribution is identical to the true
distribution: this is justified here because errors are much smaller than
measured parallaxes, so that the error function used in the deconvolution can
be assumed to be a $\delta$-function. We then extracted "measured" parallaxes
summing gaussian distributed errors (with standard deviations equal to the
errors given by HIPPARCOS) to the observed parallaxes. Finally, we estimated
the difference between the weighted average of the magnitudes for stars with
parallaxes above the threshold for both the measured sample and the original
one. We repeated the procedure 50,000 times, and we found that these "measured"
absolute magnitudes are on average 0.004~mag fainter than the original absolute
magnitudes: that is, we find a Lutz-Kelker correction of $\Delta
M_{LK}=-0.004$~mag to be appropriate for the whole sample for the given
threshold. The correction is still smaller in absolute value ($\Delta
M_{LK}=-0.002$~mag, yielding a tiny correction to ages of $\sim 0.03$~Gyr) when
only stars with $5.5<M_V<8$\ are considered. On the basis of these
considerations, we resolved on neglecting the Lutz-Kelker corrections for our
sample. Uncertainties in the distance moduli due to this neglect are less
than $0.012$~mag (in the sense that clusters may be slightly farther and
brighter). 

Lutz-Kelker corrections cannot be neglected instead for stars taken from other
samples, and the values of $M_V$'s listed in Table 1 for these stars do include
them. For the objects taken from Reid (1997), we used the Lutz-Kelker
corrections he quotes as appropriate for his sample. On the other side, it is
much more difficult to estimate the corrections for the additional stars 
included in our sample, since these were gathered from a coarse inspection of
several published lists. We finally decided rather arbitrarily to apply the
correction given by eq. (2), and appropriate for a value of $n=2.13$. This
value of $n$\ is very similar to that considered by Sandquist et al. (1996) for
their sample of subdwarfs, while it is lower than the value of $n=3.4$ obtained
by Reid (1997) for the stars of the Lowell catalogue with HIPPARCOS parallaxes.
We think a lower value of $n$\ to be more appropriate because only rather
bright stars are considered, and the value of $n$\ is expected to decrease with
decreasing limiting magnitude, due to a more severe Malmquist bias
(discriminating against distant stars). However, uncertainties related to this
assumption are well below 0.02~mag. 

It should be noted that the HIPPARCOS parallaxes for the sample of classical
subdwarfs considered by Bolte \& Hogan (1995), and VandenBerg et al. (1996) are
systematically smaller than those listed in the 1991 version of the Yale
Trigonometric Parallax Catalogue (Van Altena, Lee \& Hoffleit 1991) used by
these authors, so that the derived absolute magnitudes are on average brighter.
On average, HIPPARCOS parallaxes are smaller than those considered by
VandenBerg et al. (1996) by $7.7\pm 2.2$~mas; however, most of this difference
is due to a few discrepant cases (HD~7808, HD~84937, HD219617, and BD +11$^0$
4571, which are the stars having the largest errors in the ground-based
parallaxes). After eliminating them, the mean difference reduces to $3.6\pm
0.7$~mas (with an r.m.s. scatter of 2.7~mas, in rough agreement with the mean
quadratic error of 2.9~mas expected from errors in the ground based data
alone). There are several reasons to prefer the HIPPARCOS parallaxes. First
they were carefully tied to a uniform absolute reference system using
extragalactic objects: systematic errors are not likely to be larger than $\pm
0.2$~mas (see Kowalewski 1997). Second, the translation from relative to
absolute parallaxes is a very tricky procedure for ground-based parallaxes.
Ratnatunga \& Upgren (1997) suggest, independently of HIPPARCOS data, that the
parallaxes listed in the Yale Catalogue of Trigonometric Parallaxes are
overestimated by $3.0\pm 1.3$~mas and that errors are underestimated by $22\pm
7$\%. Their result agrees well with the difference we found for the subdwarfs.
Finally, the most recent ground-based parallaxes by Dahn (1994) are in much
better agreement with the HIPPARCOS ones than previously found. Indeed we find
that Dahn's parallaxes are only $1.1\pm 0.5$~mas larger for the 11 stars in
common between his and our lists.

The mean difference between the absolute magnitudes by VandenBerg et al. (1996)
(after application of the Lutz-Kelker corrections) and the HIPPARCOS ones is
$0.53\pm 0.14$~mag. Once the most discrepant results (HD7808, HD84937,
HD219617, BD+11$^0$ 4571, and HD19445) are excluded, the average difference
reduces to $0.24\pm 0.06$~mag. If, for ground based observations, we limit
ourselves to stars with $\Delta \pi/\pi<0.12$\, the mean difference is $0.20\pm
0.04$~mag. These comparisons directly lead to the conclusion that the distance
scale provided by the superior HIPPARCOS parallaxes is longer than that
obtained from ground-based observations. 

We anticipate that {\bf everything else being constant, globular cluster ages
derived exploiting this new distance scale are about 2.8~Gyrs younger than
those derived from ground-based parallaxes for the local subdwarfs}. 

\subsection{Colors}

While parallaxes are the basic ingredient in our analysis, they are useless if
appropriate determinations of magnitudes, colors, and chemical composition are
not available. 

Average V magnitudes and colors (Johnson $B-V$\ and $V-K$, and Str\"omgren
$b-y$, $m_1$\ and $c_1$: for brevity, only $B-V$\ values are listed in Table~1)
for the programme stars were obtained from a careful discussion of literature
data. We compiled also data for $R-I$\ and $V-I$\ colors, but decided not to
use them due to the difficulty in the definition of a uniform standard system
(uncertainties in the color transformations showed up as a large scatter in the
individual values). In our averages, we used also the $V$\ magnitudes and
$B-V$\ colors provided by the Tycho mission (Grossmann et al. 1995), after
correcting them for the systematic difference with ground-based data (the
correction is however very small, 0.003~mag in $B-V$). Complete tables with
data and details on the standardization procedure will be given in Gratton
et al. (in preparation). 

\subsection{Metallicities}

Abundances for about two thirds of the programme stars were derived from analysis
of high dispersion spectra purposely acquired for this programme using the 2.7
m telescope at McDonald and the 1.8 m telescope at Cima Ekar (Asiago). Very
high quality McDonald spectra (resolution R=70,000, $S/N\sim 200$, spectral
coverage from about 4000 to 9000 \AA) were available for 21 stars (most of them
with [Fe/H]$<-0.8$) thanks to the courtesy of dr C. Sneden; Cima Ekar telescope
provided somewhat lower quality spectra (resolution R=15,000, $S/N\sim 200$,
two spectral ranges 4,500$< \lambda < $7,000 and  5,500 $<\lambda<$to 8,000
\AA) for 65 stars. The large overlap between the two samples (15 stars) allowed
us an appropriate standardization of the equivalent widths used in the analysis.
The abundance derivation followed the same precepts of the reanalysis of $\sim
300$\ field and $\sim 150$\ GC stars described in Gratton, Carretta \& Castelli
(1997a) and Carretta \& Gratton (1997). The availability of a homogenous set of
high quality abundances (standard errors $\sim 0.07$~dex) for both field
subdwarfs and GC giants is one of the basic ingredients of our study. Details
of the abundance derivation will be given elsewhere;  here we only mention that
the analysis provided the abundances of Fe, O, and of the $\alpha-$elements Mg,
Si, Ca, and Ti. This is a crucial point, since the position of the main
sequence and the shape of the c-m diagram in the turn-off region are
significantly affected by the abundances of these elements. 

\begin{figure} 
\centerline{\hbox{\psfig{figure=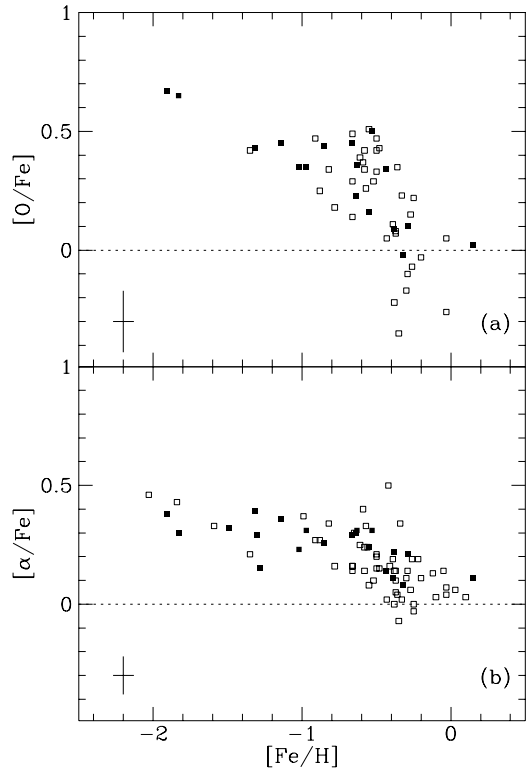,width=13.0cm,clip=}}}  
\medskip                                                                       
\caption{Runs of the overabundances of O (panel a) and $\alpha-$elements
(panel b) as a function of [Fe/H] for the programme subdwarfs. Filled
squares are abundances from McDonald spectra; open squares are abundances
from Asiago spectra. Error bars are at bottom left }
\label{f:fig2} 
\end{figure}

We found that O and the other $\alpha-$elements are overabundant in all
metal-poor stars in our sample (see panels $\it a$ and $\it b$ of
Figure~\ref{f:fig2}). The average overabundances (including also the objects 
not listed in Table~1) in stars with [Fe/H]$<-0.5$ are: 
$$ {\rm [O/Fe]}= 0.38\pm 0.13$$
$$ {\rm [\alpha/Fe]}= 0.26\pm 0.08,$$
where the given error bars are the r.m.s. scatter of individual values around
the mean, and not the standard deviation of the sample (which is 0.02~dex in
both cases). Note that O abundances were derived from the permitted IR triplet,
and include non-LTE corrections computed for each line in each star following
the precepts of Gratton et al. (1997b). \footnote{ The moderate value of the O
excess derived from the IR permitted lines is a consequence of the rather high
temperature scale adopted (see also King 1993), which directly stems from the
use of the Kurucz (1993) model atmospheres and colors. When deriving
temperatures, we applied the empirical corrections of Gratton et al. (1997b). If
this procedure is adopted, abundances from permitted OI lines agree with those
determined from the forbidden [OI] and the OH lines.} 

Several external comparisons are possible. Here we will only mention four
cases, which are of particular relevance due to the high quality of the
spectroscopic data and of the analysis. 

Very recently Balachandran \& Carney (1996) found [Fe/H]=$-1.22\pm 0.04$\ and
[O/Fe]=$+0.29\pm 0.05$\ for HD103095; our values are [Fe/H]=$-1.24\pm 0.07$,
[O/Fe]=$+0.37\pm 0.08$, and [$\alpha$/Fe]=$+0.23\pm 0.08$. The agreement is of
course excellent. 

We have analyzed six of the  stars in the Edvardsson et al. (1993) sample. On
average, abundance residuals (our analysis$-$Edvardsson et al) are $+0.08\pm
0.03$, $-0.02\pm 0.03$, and $+0.02\pm 0.02$~dex for [Fe/H], [O/Fe], and
[$\alpha$/Fe] respectively. The agreement is good, and the star-to-star scatter
is very low ($\sim 0.05$~dex); the residual differences are mainly due to our
higher temperature scale. 

We have six stars in common also with of Tomkin et al. (1992; hereinafter
TLLS). The parameter range is large; furthermore, TLLS had a much more
restricted wavelength coverage. This translates into a larger star-to-star
scatter. The average differences (our analysis$-$TLLS) are also large:
$+0.34\pm 0.04$\ and $-0.31\pm 0.07$~dex for [Fe/H] and [O/Fe] respectively.
These differences are due to different assumption in the analysis: first we
adopted a higher temperature scale (on average, our $T_{\rm eff}$'s are larger
by $136\pm 28$~K); second, our abundances are computed using a solar model
extracted from the same grid used for the programme stars, while TLLS used the
Holweger \& M\"uller (1974) empirical model for the Sun, and flux-constant
MARCS models (Gustafsson et al. 1975) for the programme stars; third, our
non-LTE corrections to the O abundances are slightly larger than those used by
TLLS. If the same analysis procedure is adopted, TLLS $EW$s give abundances
almost coincident with the present ones. 

Finally, Gratton et al. (1997a; 1997c) gave a homogenous reanalysis of the
original $EW$s for about 300 metal-poor field stars. On average, present Fe
abundances are larger by $0.02\pm 0.02$~dex (11 stars, r.m.s.
scatter=0.06~dex), while Gratton et al. (1997c) found an average overabundance
of [O/Fe]=$0.45\pm 0.02$\ in metal-poor stars. Since the same analysis
procedure was adopted, these difference are entirely due to random errors in
the $EW$s and small differences in the adopted colors. In the following, we
will assume that Gratton et al. abundances are on the same scale of the present
analysis. Note that Gratton et al. considered some of the HIPPARCOS programme
stars for which no high dispersion spectra were analyzed here. 

Inspection of Figure~2 might suggests the presence of a trend of [O/Fe] with
[Fe/H] (the trend is not as obvious for [$\alpha$/Fe]). However this result is
based on very few stars; the much more extensive analysis by Gratton et al.
(1997c), which is completely consistent with the present one, does not support
the presence of such a trend. It is also interesting to note that very recently
King (1997) obtained a rather small excess of O and $\alpha-$elements for the
proper motion pair HD134439-134440. He suggests that this is an evidence for
inhomogeneities in the halo, possibly related to accretion events. Our 
$\alpha-$element abundances for these stars are also a little lower than
average (although somewhat larger than those found by King): it may well be
that some intrinsic scatter does exist in the element-to-element ratios amongst
halo stars. However, this scatter is not much larger than observational errors,
and King results suggest that a moderate excess of $\alpha-$elements is present
even in HD134439-134440. Note also that the overabundance of O and $\alpha-$
elements found for the field subdwarfs is similar to the excesses found for GC
giants (apart from those stars affected by the O-Na anticorrelation, see Kraft
1994). In the following we will assume that O and $\alpha-$elements are
uniformly overabundant by 0.3~dex in both field and cluster stars with
[Fe/H]$<-0.5$. 

For the stars (mostly rather metal rich) lacking high dispersion analysis,
abundances were obtained from the Str\"omgren photometry using the calibration
by Schuster \& Nissen (1989), and from the empirical calibration of the cross
correlation dips by Carney et al. (1994). We found that these abundances
correlate very well with high dispersion results, apart form zero-point
offsets. \footnote{ These offsets are mainly due to different assumptions about
the solar abundances in the high dispersion analyses originally used in the
calibrations. The solar abundances were not derived homogeneously with the
stellar abundances in most abundance analysis made before the Kurucz (1993)
models became available. This problem is now solved by the ability of Kurucz
models to adequately reproduce most of the solar observations to the level of
precision of the present analysis.} We then corrected the former to place them
on our scale. Errors (derived from the r.m.s. scatter of differences with our
estimates) are 0.13~dex for abundances from Str\"omgren photometry and 0.16~dex
for those derived from Carney et al. (1994). For those stars having both
(independent) estimates errors are 0.09~dex. 

\subsection{Binaries}

\begin{figure} 
\centerline{\hbox{\psfig{figure=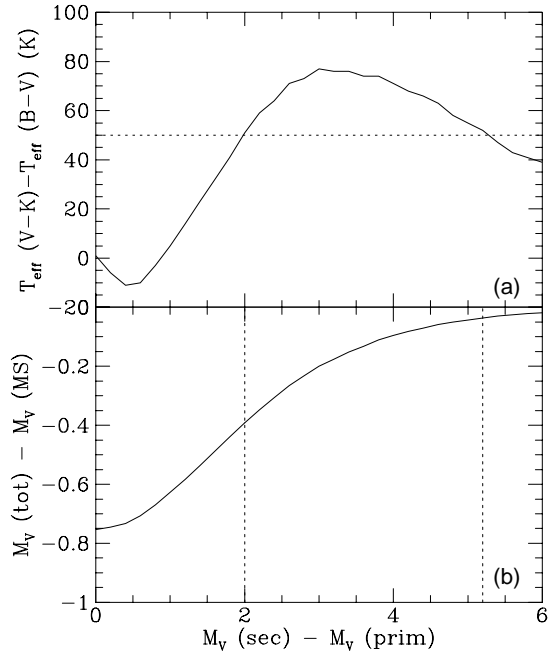,width=13.0cm,clip=}}}  
\medskip
\caption{(a) Run of the difference in  effective temperatures derived
from $V-K$\ and $B-V$\ colors as a function of the magnitude difference
between primary and secondary component. This method may be used to detect companions 2
to 5 mag fainter than the primary if a threshold value of 50~K is adopted
(dashed line). (b) Run of the difference between the total system $M_V$\ and
the $M_V$\ of the main sequence at the same $B-V$\ color as a function of the
magnitude difference between primary and secondary. The dashed line limits the
region where the IR excess method can be used to detect binaries using the
threshold of 50~K}
\label{f:figadd1} 
\end{figure}

\begin{figure} 
\centerline{\hbox{\psfig{figure=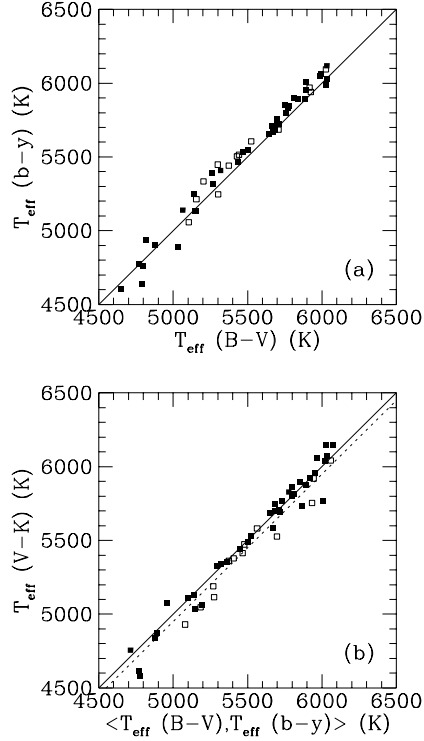,width=13.0cm,clip=}}}  
\medskip
\caption{ (a) Comparison between effective temperatures derived from $B-V$\ and
$b-y$\ colors. Open squares are known or suspected binaries from radial
velocity variations or from astrometry; filled squares are supposed "single"
stars. The solid line represents identity between the two temperatures.
Binaries and "single" stars are almost indistinguishable in this diagram. (b)
Comparison between average effective temperatures obtained from the blue colors
($B-V$ and $b-y$) and those derived from $V-K$. The dashed line is offset by
50~K from the identity line: stars below it are suspected to be binaries due to
the presence of an IR excess } 
\label{f:figadd2} 
\end{figure}

One of the largest sources of systematic errors in our derivation of the
position of the main sequence as a function of metallicity is the
contamination of the sample of main sequence stars by unresolved binaries. In
order to clean up our sample from binaries we considered four different sets of
information: 
\begin{itemize}
\item A rather large fraction of the programme stars are known or
suspected spectroscopic binaries from the long term, high precision, monitoring
programmes of Carney et al. (1994), and Stryker et al. (1985). Three stars were
recognized as doubled-lined spectroscopic binaries (SB2) on the basis of our
spectra (two of them not previously known as SB2). Out of the 34 stars listed
in Table 1, 9 are known and 2 are suspected spectroscopic binaries, namely
HD111515 and HD103095(=Gmb~1830, the only stars with [Fe/H]$<-1$\ within 10 pc
from the Sun). Data for these two last stars are from Carney et al. (1994).
While the r.m.s scatter for HD111515 (1.0~km~s$^{-1}$\ from 6 observations) is
well above the average value for the metallicity of this stars (supporting the
identification of the star as a binary), the evidence of binarity is less
significant for HD103095. Carney et al. (1994) obtained a r.m.s. scatter of
only 0.6~km~s$^{-1}$\ from 212 observations of the star spread over 12 yrs.
Moreover, none of the methods listed below flags this star as a binary (in
particular, there is no indication of an IR excess signaling a secondary
component from 2 to 5 magnitudes fainter than the primary). In the following we
will then assume that HD103095 is a single star. 
\item Three additional programme stars displayed rotationally broadened lines,
perhaps due to tidal locking in close binary systems. None of these
stars was included in Table 1, but one (HD7808) is in the list considered by
VandenBerg et al. (1996).
\item HIPPARCOS provides information about astrometric binaries. Two of the
stars of Table 1 (G246-038 and HD224930) are classified as suspected
astrometric binaries. 
\item Finally, some binaries could be discovered due to an overall IR excess
caused by a cooler secondary component. This method is very powerful since it
does not depend on the angle of view or on the orbital motion, and it is
capable to detect binaries where rather large magnitude differences occur
between the two components. These are the most dangerous systems for systematic
errors because they do not stand out obviously in other plots.  Panel {\it a}
of Figure~\ref{f:figadd1} shows the difference between temperatures derived
from the $V-K$\ and the $B-V$\ color respectively, as a function of the
difference in magnitude between primary and secondary component of the system
(we assumed $M_V(\rm primary)=+6$\ and [Fe/H]=$-1$\ for this particular
example). Rather large differences ($>50$~K) are found when the secondary is
from 2 to 5 mag fainter than the primary. In panel {\it b} we have plotted the
difference between the magnitude of the system and that of the main sequence at
the same total $B-V$\ color (this is the error we would introduce when deriving
the distance moduli of globular clusters). This figure shows that even rather
faint secondaries may cause significant errors in the position of the main
sequence, and that the range where the IR excess method efficiently flags
binaries corresponds to where the stellar luminosity would be overestimated
from 0.05 to 0.4~mag if binarity is neglected. 

Effective temperatures for the stars in our sample derived from the blue colors
$B-V$\ and $b-y$, are compared in panel {\it a}) of Figure~\ref{f:figadd2}.
Binaries (open symbols) and supposed single stars (filled squares) are almost
indistinguishable in this diagram. In panel {\it b} the average effective
temperatures from $B-V$\ and $b-y$, are compared  with those derived from the
$V-K$. Although there is some observational scatter, we find that temperatures
derived from $V-K$\ for the known binaries (open symbols) are on average lower
than those derived from blue colors, while this does not occur for the majority
of "single" stars. However, there are a few additional "single" stars for which
the temperature derived from $V-K$ is more than 50~K below than that derived
from the blue colors: these stars are suspected to be binaries. Two of them
(G125-04 and G140-46) are included in Table~1. 
\end{itemize}

Summarizing, there are 10 known and 4 suspected binaries among the 34 stars
listed in Table 1. All these stars were flagged in the Note column of Table~1:
SO are spectroscopic binaries with orbital solution; SB are spectroscopic
binaries without orbit determined; S? are suspected spectroscopic binaries; AB
are stars suspected to be binaries from the HIPPARCOS astrometric solution; IR
are stars suspected to be binaries due to the presence of an IR excess. All
these stars were not used in the MS-fits since both their magnitudes and colors
may be significantly distorted by the presence of a companion. When considering
the total binary fraction of our sample, we should add to these stars the
common proper motion pair HD134439-HD134440, for which the separation is so
large that does not disturb our analysis, and was then kept. The overall
fraction of binary systems is then about half of the sample, similar to typical
values for nearby population I stars. 

Thirteen of the 20 supposed {\it bona fide} single stars listed in Table 1 have
$M_V>5.5$ and will then be used in the MS-fits; 2 of these stars have a
metallicity [Fe/H]$<-1.5$, 7 are in the range $-1.5<$[Fe/H]$<-1.0$, and 4 are
in the range $-1.0<$[Fe/H]$<-0.4$. In Section 5.1 we will examine the errors in
our distance moduli for GCs due to possible binary contamination in this
residual sample of {\it bona fide} single stars. 

\section{LOCATION OF THE MAIN SEQUENCE AS A FUNCTION OF METALLICITY}

Our sample of field stars with accurate absolute magnitudes and metal
abundances was used to determine the {\it absolute} location of the main
sequence as a function of metallicity. This step is basic for the fitting of
the main sequence and the derivation of cluster distances. It further provides
a first important test on theoretical isochrones. 

In principle the procedure simply consists in selecting stars in a given range
of metal abundances. However, given the rather large sensitivity of the main
sequence colors to metallicity, our sample is not numerous enough for this
direct application, since the metallicity range must be selected to be rather
broad in order to include a large enough set of stars. 

To reduce random and systematic errors due to the difference between cluster
and field star metallicities, colors of the field stars must be corrected
for the corresponding shift of the main sequences. While these offsets are
quite small, uncertainties can be further reduced by an appropriate selection
of the corrections. In principle, these can be determined {\it empirically}
using colors and metallicities of the subdwarfs themselves, once the slope of
the main sequence (which is nearly independent of metal abundance) is properly
taken into account. 

Our procedure consisted in determining the color of MS stars at an absolute
magnitude of $M_V=+6$\ (this value was selected in order to be typical of
unevolved stars in our sample). Colors for stars with absolute magnitude
$5.5<M_V<8$\ were corrected to those they would have for $M_V=+6$\ using the
slope of the main sequence. Evolutionary effects cannot be neglected for stars
brighter than $M_V=5.5$, and possible age differences between field stars and
globular clusters may cause systematic errors in the distance derivations: 
for instance if a field star is 2~Gyr younger than a GC having the same 
metal abundance, it would lie below the cluster main sequence by only 0.01~mag
if its absolute magnitude is $M_V=5.5$, but by as much as 0.05~mag if the
absolute magnitude is $M_V=5.0$\ (this difference increases rapidly with
decreasing absolute magnitude for $M_V<5$). On the other side, atmospheres of
stars fainter than $M_V=8$\ are dominated by molecular bands, so that
metallicities cannot be reliably determined. 

In practice, these corrections can be determined either empirically,
from the slope of the GC main sequences (if some preliminary guess of the
distance modulus is available), or from theoretical isochrones. Note that the
slope is not constant over the whole magnitude range here considered: hence
we considered two distinct values for the magnitude range $5.5<M_V<6$\ and
$6<M_V<8$.

The values of the MS slope we adopted were the average of those determined from
the mean loci of the globular clusters we will consider in Sect. 5. These
averages were $\Delta (B-V)/\Delta  M_V=0.168\pm 0.004$ for the range
$5.5<M_V<6$\, and  $\Delta (B-V)/\Delta M_V=0.216\pm 0.007$ for the range
$6<M_V<8$\ (the error bar is the standard deviation of the mean of 10
photometries for 9 clusters). For comparison, the slopes given by theoretical
isochrones in the range $6<M_V<8$\ are $\Delta (B-V)/\Delta  M_V=0.22$\ when
using VandenBerg (1992) transformations, and $\Delta (B-V)/\Delta M_V=0.18$\
when using the Kurucz (1993) transformations. While the results are slightly
different from the last theoretical prediction, our main conclusions do not
depend on this choice. 

\begin{figure} 
\centerline{\hbox{\psfig{figure=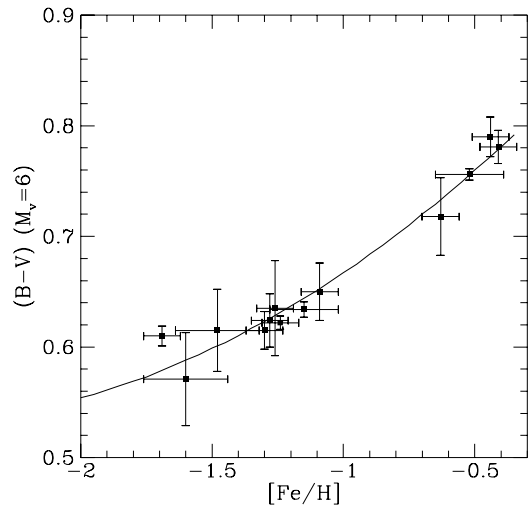,width=13.0cm,clip=}}}  
\medskip                                                                       
\caption{ Run of the $B-V$\ color for unevolved main sequence stars (at $M_V=+6$)
as a function of metallicity [Fe/H]. Only {\it bona fide} single stars with 
$5.5<M_V<8$\ are plotted. The relation used in this paper is overimposed } 
\label{f:fig3} 
\end{figure}

Figure~\ref{f:fig3} displays the run of the $B-V$\ color for unevolved main
sequence stars (at $M_V=+6$) as a function of metallicity [Fe/H]. 
The unweighted best fit relation is:
\begin{equation}
(B-V)_{Mv=6} = 0.915 + 0.356 {\rm [Fe/H]} + 0.099 {\rm [Fe/H]}^2
\end{equation}
This relation has a minimum at [Fe/H]$\sim -1.8$, at odds with theoretical
expectations: the relation between color and metallicity is in fact monotonic
for all isochrones sets we considered, a result which is expected to be robust
(VandenBerg, private communication). Given this discrepancy, we preferred to
adopt a semiempirical approach; we find that a reasonable representation of the
color-metallicity dependence shown in Figure~\ref{f:fig3} is given by the
relation: 
\begin{equation}
(B-V)_{Mv=6} = 0.876 + 0.257 {\rm [Fe/H]} + 0.048 {\rm [Fe/H]}^2
\end{equation}
that is obtained by using various sets of isochrones (e.g. Straniero \& Chieffi
1997; D'Antona, Caloi \& Mazzitelli 1997; Bertelli et al. 1997) and VandenBerg
(1992) transformations. The slope derived using these same isochrones and
Kurucz (1993) transformations is steeper, in worst agreement with observations.
We will then assume that eq. (4) gives the correct color-metallicity relation.
If the best fit relation had been used instead, ages younger by 2.1,
0.7, and 0.4 Gyr, would have been derived for the most metal-poor clusters
M92, M68, and M30, while ages for the other clusters would be nearly unchanged.
The average age for the six {\it bona fide} old cluster considered in Section 6
would be about 0.4 Gyr lower if the adopted relation between the colour
of the main sequence (eqt. 4), is replaced by the best fit relation of
eqt. (3).

\begin{figure} 
\centerline{\hbox{\psfig{figure=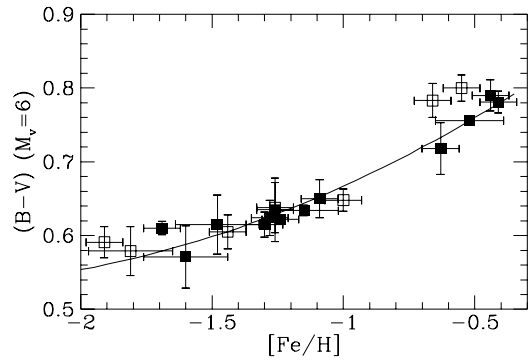,width=13.0cm,clip=}}}  
\medskip
\caption{ Run of the $B-V$\ color for unevolved main sequence stars (at
$M_V=+6$) as a function of metallicity [Fe/H]. Only {\it bona fide} single
stars are plotted: filled squares are stars in the unevolved section of the
main sequence $5.5<M_V<8$; open squares are slightly brighter stars
($5<M_V<5.5$). The relation used in this paper is overimposed } 
\label{f:fig4add} 
\end{figure}

While we stress again the importance to estimate distances by fitting only the
portion of MS not affected by evolution, we note that eq. (4) fits well also
the location of the stars that have an absolute magnitude in the range
$5.0<M_V<5.5$. This
is illustrated by Figure~\ref{f:fig4add} where we display the
$(B-V)(M_V=6)-$[Fe/H] diagram for all {\it bona fide} single stars listed in
Table~1 (an average slope of $\Delta (B-V)/\Delta M_V=0.132\pm 0.004$\ was
adopted for the range $5.0<M_V<5.5$; although in this luminosity range the
slope of the MS is weakly dependent on metallicity, errors due this dependence
are $<0.005$~mag). The larger number of metal-poor stars considered in
Figure~\ref{f:fig4add} strengthen our finding of a relatively weak dependence
of MS colors on metallicity for stars with [Fe/H]$<-1$; a similar result is
obtained by Reid (1997). 

\section{BASIC DATA FOR THE GLOBULAR CLUSTERS}

\subsection{Cluster Photometries}

Once the template main sequences have been determined, we may compare them with
the location of the GC main sequences. Given the rather steep slope of the main
sequences in the $M_V-(B-V)$\ plane (from 5 to 7, depending on luminosity), and
the rather strong dependence on metallicity, dereddened colors and
metallicities should have errors $\leq 0.01$~mag and $\leq 0.1$~dex\footnote{
Metallicities affect age derivations in two ways: (i) there is a metallicity
term in the relation between the turn-off magnitude and age ($\log t\sim 0.12
{\rm [Fe/H]}$, the exact value depending on the isochrone set); and (ii) the
distance derivation is influenced because the colors of the main sequence
depend on metallicity ($(B-V)\sim 0.12 {\rm [Fe/H]}$\ over the range
$-2<$[Fe/H]$<-1$)} in order to derive ages with an accuracy of $\sim 1$~Gyr
(the final aim of the present analysis). Clearly, the comparisons are
meaningful only if (i) photometric calibrations are very accurate, (ii)
reddenings are accurately estimated, and (iii) metallicities are both accurate
and on a scale consistent with that used for the field stars. 

As for the photometric calibrations, we generally relied upon the quality of the
original colors. By itself, such a good quality is not an obvious issue even
for very deep photometric studies (see e.g. the 0.04~mag difference between the
color of the M92 main sequence by Heasley \& Christian 1991 and that by
Stetson \& Harris 1988: this difference would translate into a 3 to 4~Gyr
difference in the age derived for this very important cluster). We hope that
the use of a not too restricted sample of carefully selected clusters may lead
to average photometric errors, and to show up anomalous cases. In the following,
we will consider three such cases (M30, NGC6752, and NGC362). In the 
discussion, we will make use of the metallicities, reddenings, and distance
moduli derived in our analysis (that is, we will use arguments based on the
self-consistency of our results). 

\begin{figure} 
\centerline{\hbox{\psfig{figure=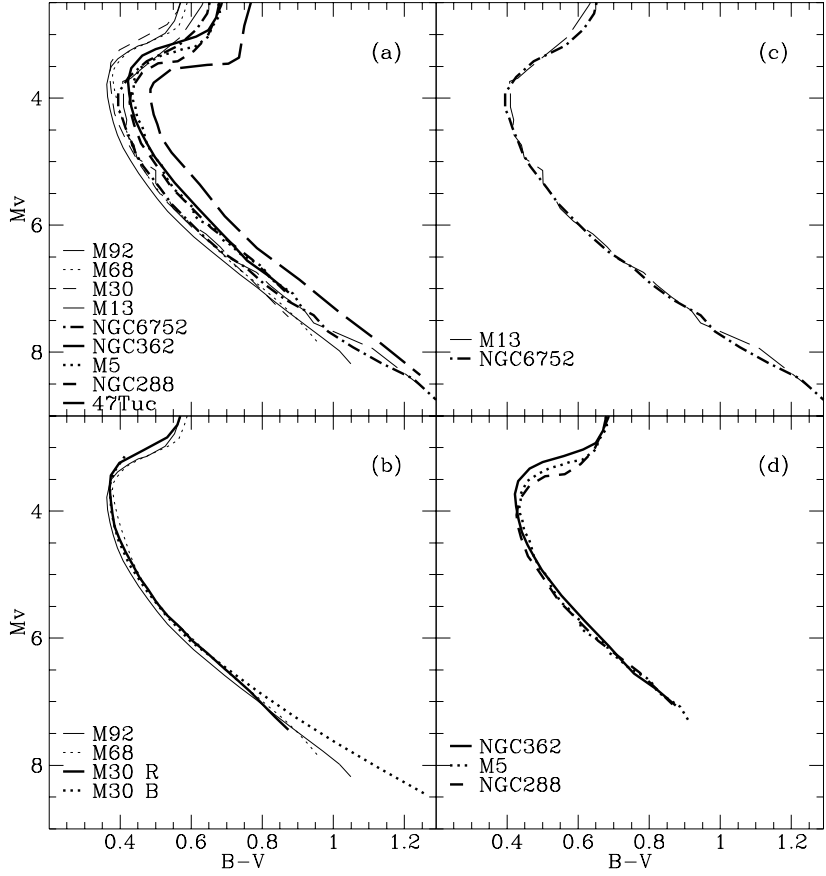,width=13.0cm,clip=}}}  
\medskip                                                                       
\caption{ (a) Fiducial mean loci of the GCs considered in this paper. For M30,
the sequence by Bolte (1987b) is shown; the cluster sequences were corrected
for the reddening and apparent distance moduli determined in the present paper.
(b) The same of panel (a), but only for the Oosterhoff II clusters M92, M68, and
M30 (for this last cluster, both the Bolte (1987b) and the Richer et al. (1988)
sequences are shown. (c) The same of panel (a), but only for the clusters with
blue horizontal branches M13 and NGC6752; the subgiant sequence for M13 is from
Sandage (1970). (d) The same of panel (a), but only for the clusters with
[Fe/H]$\sim -1.1$\ NGC362, M5, and NGC288. } 
\label{f:fig5} 
\end{figure}

Two deep main sequence photometries are available for M30 (Bolte 1987b, and
Richer et al. 1988). They agree very well for $V<21.5$, but there are large
differences at fainter magnitudes. When we compare the mean loci of M30 with
those of other metal-poor clusters (M92 and M68), the Bolte MS seems too red,
and the Richer et al. too blue below $V=21.5$\ (see panel {\it b} of
Figure~\ref{f:fig5}). We have adopted the simple average of the distance moduli
obtained from the two individual fiducial lines. 

Penny \& Dickens (1986) photometry for NGC6752 was obtained using very wide
filters, and then transformed to the Johnson $BV$\ system. VandenBerg et al.
(1990) pointed out that Penny \& Dickens photometry on the lower main sequence
differs substantially from that obtained by Buonanno et al. (1986b). Stetson \&
Harris (1988) observed that (using reddening and distance moduli different from
those adopted here) the lower main sequence of Penny \& Dickens passes over the
red of 47 Tuc, a cluster with a higher metallicity, suggesting an error in
Penny \& Dickens photometry. VandenBerg et al. (1990) proposed a mean loci
sequence bluer by 0.03 mag at $V\sim 19.75$\ ($M_V=6.58$\ using our distance
modulus): no value is given for fainter magnitudes. With this correction, the
main sequence of NGC6752 is almost indistinguishable from that of M13, a
cluster with a very similar metallicity and HB morphology (see panel c of
Figure~\ref{f:fig5}). On the basis of this argument, we have hereinafter
adopted the correction proposed by VandenBerg et al. (1990). 

In the case of NGC~362, VandenBerg et al. (1990) proposed a fiducial main
sequence that is 0.03 mag redder than that originally found by Bolte (1987a).
The reason for this difference is not entirely clear; however Bolte claimed an
accuracy of only 0.02~mag in his photometric zero points. Adoption of Bolte's
fiducial line yields a horizontal branch very bright for the metallicity of 
NGC~362, a result difficult to understand. On the other side, both the main
sequence and the subgiant branches of NGC362 would nicely overlap the sequence
for NGC288 and M5 (two clusters having a similar metallicity)
if VandenBerg et al. (1990) fiducial line (and our estimate of the distance
modulus) are adopted (see panel {\it d} of Figure~\ref{f:fig5}). Hence, also in this
case we adopted the correction proposed by VandenBerg et al. (1990). 

Finally, it should be noted that the cluster mean loci are usually defined
using an algorithm which identifies the modal value of the color distribution
of the MS stars in each magnitude bin (see e.g. Sandquist et al. 1996). In this
way, the authors try to correct for the systematic reddening of the main
sequence due to the contamination by unresolved pairs (either real binaries or
apparent couples). The net effect of a fiducial line drawn too red is an
underestimate of the cluster distance derived from subdwarf fitting, and an
overestimate of its age. Clearly use of a modal value should be preferred to
simple averages; however, even the observed modal value may be slightly redder
than the true single star sequence, due to the presence of very faint
contaminants causing a broadening of the main sequence comparable to the
observational errors (a few hundreths of a magnitude over the range of interest
in the best MS photometries). The difference between the modal value and the
true value is smaller than this, depending on the internal quality of the
photometry, the degree of crowding of the field, and the magnitude and color
distribution of the contaminants. A thorough discussion of each of the
photometric study considered here is clearly beyond the purposes of the present
paper. Hereinafter we will neglect this source of systematic errors. 

\subsection{Interstellar reddening}

Uncertainties in the values commonly adopted for the interstellar reddenings
of globular clusters may well be as large as $\sim 0.05$~mag, implying
errors as large as 5~Gyr in the derived ages ($\sim 8$~Gyr for the most
metal-rich ones). With such large uncertainties the whole main sequence
fitting procedure would be nearly useless. To reduce this concern we have 
adopted the following procedure :

\begin{itemize}

\item Only clusters projected toward directions of very low interstellar
absorption were considered. In practice, we selected only those nine clusters
for which $E(B-V)\leq 0.05$ according to Zinn (1980), the metallicity is
determined from high dispersion spectra of individual red giants on the
same scale adopted for the field subdwarfs (Carretta \& Gratton 1997),
and deep (and hopefully accurately calibrated) MS photometry was available.

\item However, even for these best clusters the error bars of Zinn (1980)
reddening estimates (often used for these comparisons: 0.02 to 0.03~mag) were
deemed too large for the present purposes. We then decided to average Zinn's
values (themselves an average of values available in the literature at that
epoch) with other independent estimates. A large, homogenous set of reddening
estimates has been provided by Reed, Hesser \& Shawl (1988), based on
integrated photometry of globular clusters. Whenever possible, we added to
these two sets new reddening estimates based on Str\"omgren and $H_\beta$
photometry of B-F stars projected on the sky within 2~degrees from the clusters
(since all these clusters lie at high galactic latitudes, reddening is not
expected to change too rapidly with the position on the sky). To this purpose,
we used the calibrations by Crawford (1975, 1978, 1979): essentially we
compared the observed (reddened) $b-y$\ index with the value predicted from
$H_\beta$, $m_1$, and $c_1$. The field stars having Str\"omgren photometry are
closer than the GCs; to avoid underestimates of the cluster reddenings, we only
considered stars at distances above 120~pc from the galactic plane : this value
is larger than the scale height of interstellar dust at the solar circle as
determined from CO and IRAS dust emissivities (Gaussian dispersions of 50 and
100~pc respectively: Burton 1992). The Str\"omgren photometry is in principle
able to provide very precise reddening estimates: standard errors for
individual stars, as determined from the star-to-star scatter is only $\Delta
E(B-V)=0.024$~mag. Unfortunately, adequate photometry exists only for the
southern clusters, and in several cases (like e.g. M92) we were unable to find
any suitable star in the Str\"omgren photometry catalogue (Hauck \& Mermilliod
1990). Furthermore, the reddening value for the best determined cluster (NGC
288, 7 stars) came out negative ($E(b-y)=-0.017$~mag), and all values are
systematically smaller than those published by both Zinn (1980) and Reed et al.
(1988). This may be due to systematic errors in the adopted calibrations. We
then corrected these reddening estimates upwards to put them on the same scale
provided in the two other papers. The individual reddening estimates are listed
in columns 4 to 6 of Table~2. 

\item Final adopted values for the reddening were then the simple mean of: (i)
the estimates based on Str\"omgren photometry of nearby field stars, (ii) the
Zinn (1980) values, and of (iii) the estimates by Reed et al. (1988), (see
Column 7 of Table~2). These reddenings have mean quadratic errors of 0.008~mag
(standard deviation of the mean), and represent the best determinations for
these nine clusters up-to-date. However, some uncertainty ($\sim 0.01$ mag)
still exists on their zero-point. As a comparison, we note that our reddening
value for M92 ($E(B-V)=0.025\pm 0.005$) is somewhat higher (though within the
error bar) than that adopted by VandenBerg et al. ($E(B-V)$=0.02). Everything
else being constant, this implies a distance modulus 0.04~mag larger and, in
turn, an age 0.5 Gyr younger for this very important cluster. 
\end{itemize}

\subsection{Cluster metal abundances}

The final input parameter used in the main sequence fitting procedure is the
metal abundance of the globular clusters. The most accurate cluster metallicity
determinations are those from high dispersion spectroscopy of individual
giants. For consistency with the field subdwarf (and the solar) analysis we
adopted the very recent determinations by Carretta \& Gratton (1997). Small
corrections were applied, though, to account for the difference between the
reddening values adopted in Carretta \& Gratton and the new values found in
this paper. Carretta \& Gratton analysis of the GC giants strictly follows the
procedure adopted here for the field dwarfs. On the other side, the atmosphere
of giant stars is considerably different from that of dwarfs, and it may be
possible that some difference in the metallicity scales still exists.
Unfortunately, there is at present no metal abundance derivation for GC dwarfs
from high dispersion spectra (Li abundances have been determined for a few
stars near the turn-off of NGC6397 and M92 by Pasquini \& Molaro 1996, and
Boesgaard 1996). There are however a number of circumstantial evidences
supporting the consistency of dwarf and giant abundance scales: (i) the data of
individual stars in Carretta \& Gratton, show no obvious trend of [Fe/H] with
luminosity over more than 3 mag along the upper giant branch of GCs; (ii) there
is a fairly good agreement between abundances obtained for red giants and
horizontal branch stars in M4 (see e.g. Clementini et al. 1994); and (iii)
Folgheraiter et al. (1995) obtained [Fe/H]=$-1.4$\ from Str\"omgren photometry
of main sequence stars in NGC~6752, in good agreement with our value of
[Fe/H]=$-1.43$\ (although the Folgheraiter et al. result is quite uncertain due
to inadequate calibration of the photometry). Clearly, high dispersion analysis
of GC dwarfs would be highly welcomed. In the following discussion, we will
assume that cluster and subdwarf metallicity scales do not differ by more than
0.1 dex. 

As a reference, we point out that Carretta \& Gratton abundances for metal-poor
and metal-rich clusters agree quite well with Zinn \& West (1984) values
(usually adopted for globular clusters), while abundances are systematically
larger for clusters of intermediate metal abundance (like M13) by as much as
0.2--0.3~dex. \footnote{ Carretta \& Gratton indeed found a very good quadratic
correlation between high dispersion and Zinn \& West's abundances.} 

{\bf Such a difference in metallicity produces distance moduli larger by $\sim
0.15$~mag, and, in turn, ages younger by $\sim 2$~Gyr for these clusters}. On
the other side, our metal abundance for the metal-poor cluster M92
([Fe/H]=$-2.15$) is slightly smaller than the value used by VandenBerg et al.
([Fe/H]=$-2.14$), we should then derive an age 0.1 Gyr larger for this cluster.

The final adopted metallicities have internal errors of $\pm 0.06$~dex; this
yields a random error of 1.6\% in the age, and errors in the distances varying
from 0.02 to 0.07~mag (and corresponding uncertainties of 4\% and 10\% in ages,
respectively) with [Fe/H] raising from $-2.0$\ to $-0.7$. 

\section{GLOBULAR CLUSTERS DISTANCES}

Once template main sequences for the appropriate metallicity are determined,
and cluster reddenings are known, cluster distance moduli may be derived by
directly comparing the apparent magnitude of the cluster main sequence at a
given color and the absolute magnitude of the template main sequence at that
color. 

In practice, we fitted the main sequence loci with the absolute magnitude of
the subdwarfs used to define the template main sequence; the distance modulus
was determined by a least square procedure. The weights include the errors in
the parallaxes, photometry and metal abundances for each star (see Table~1).
Only single stars with $M_V>+5.5$\ are considered here, since brighter objects
are evolved off the zero age main sequence and the derived distances might be
affected by age differences between cluster and field stars. 

In order to reduce any concern in the metallicity-color transformation for MS
stars, we only considered subdwarfs in a restricted metal abundance range
around the metallicity of the globular cluster. Given the narrow ranges
adopted, color changes due to this transformation are below 0.06~mag in all
cases except for the most metal rich cluster (47~Tuc), where they may be as
large as 0.10 mag (due to the increased steepness of the metallicity color
relation). Only stars having colors overlapping those of the cluster fiducial
mean sequences were considered. 

Column 6 of Table~3 gives the derived distance moduli. Metallicity ranges
adopted for the reference subdwarfs are also shown. Standard deviations of the
weighted mean of the distance moduli determined from the subdwarfs range from
0.06 mag (for the three most metal-poor clusters) to 0.04 mag (for the other
clusters). However, errors in the distance moduli should take into account
errors in the colors of the GC mean loci due to reddening (see Table 2) and
photometry (we arbitrarily assumed error bars of $\sigma (B-V)=0.01$\ for all
clusters), as well as the uncertainties in the cluster metallicities (0.06~dex;
the effect on colors ranges from $\sigma (B-V)=0.003$\ to 0.012, depending on
the metallicity). All these sources of error were included in the error bars
attached to the distance moduli listed in Table~3: typical errors are 0.08 mag.
Systematic errors (due e.g. to residual systematic differences between the
cluster and subdwarf metallicity scale) should be of the same order of
magnitude. 

\begin{figure} 
\centerline{\hbox{\psfig{figure=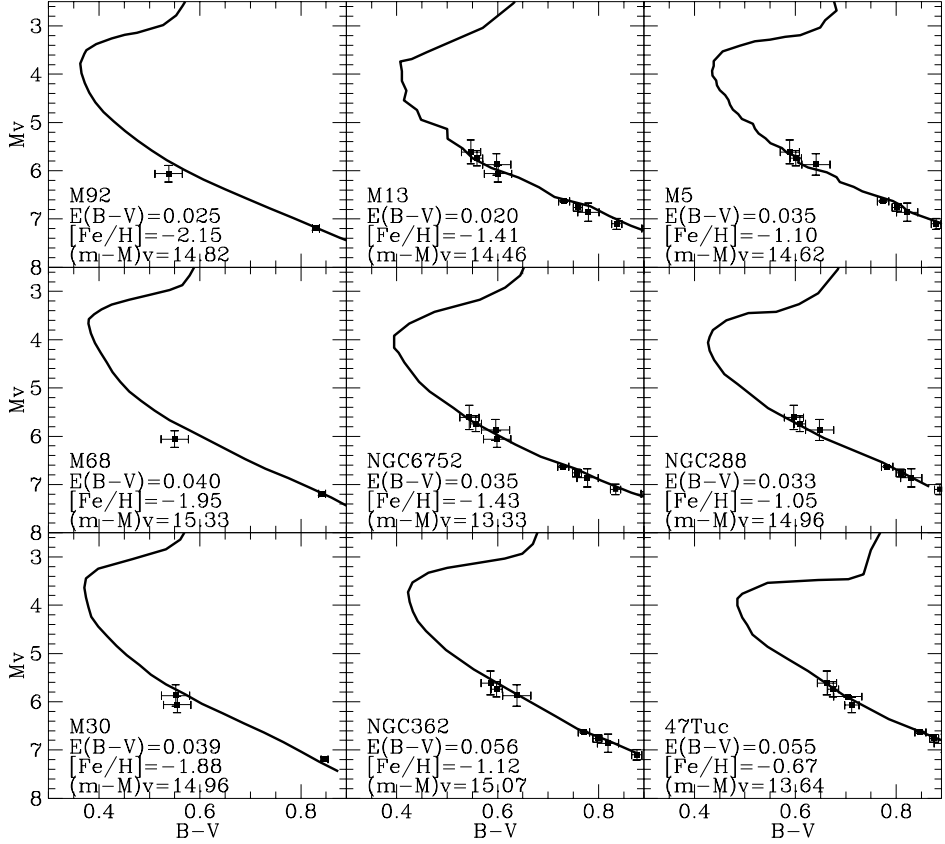,width=13.0cm,clip=}}}  
\medskip                                                                       
\caption{ Fits of the fiducial mean loci of the GCs considered in this paper
with the position of the subdwarfs of Table 1 (only {\it bona fide} single
stars with $M_V>5.5$\ are shown). The values of the parameters adopted in the
present analysis are shown in each panel. The fiducial sequence of M30 
is from Bolte (1987b)} 
\label{f:fig4} 
\end{figure}

Figure~\ref{f:fig4} displays the fits of the individual GC main sequences with
the nearby subdwarfs of the proper metallicity. The scatter of individual
points generally agrees with the expected error bars. 

\subsection{Errors in distance moduli due to undetected binaries}

As mentioned in Section 2.4, the possible contamination of our sample of
metal-poor subdwarfs by undetected binaries is one of the major concern in our
distance derivations. In this subsection we provide an estimate of the average
correction to be applied to our data to account for this effect, as well as an
estimate for the related uncertainty. 

The binary correction may be written as the product of two terms: the average
offset $X$\ for each binary, and the probability $p_s$\ that a {\it bona fide}
single star is actually a binary. In principle, both these quantities can be
derived {\it a priori}. The average offset $X$\ is a weighted average of the
magnitude offsets of Figure~\ref{f:figadd1} (although these refer to a
particular value of $M_V$\ and metallicity, they can be assumed to be
representative of all stars in our sample), where the weights take into account
the luminosity function of the secondary components. 
As discussed by Kroupa, Tout \& Gilmore
(1993), this quantity is largely uncertain. Observations of the
distribution of binaries in clusters (open as well as globular) are not very
useful here, since the distribution of the secondary components for the field
binaries is expected to be different from that of open cluster binaries, where
mass segregation increases the fraction of nearly equal mass systems in the
central regions (from which data about binary distribution are drawn); and the
present population of binaries in globular clusters is not primordial as binary
systems are likely generated by the cluster dynamical evolution. To provide a
first guess for $X$, we followed the discussion in Kroupa et al., and
assumed that the mass of the secondaries is not related to that of primaries
for low mass field stars ($M<0.8~M_\odot$), this point is
still controversial though (see e.g. Duquennoy \& Mayor 1991). We farther 
assumed that the
luminosity function of the secondaries is simply the luminosity function of the
field stars fainter than the primary component (since we are considering
unevolved stars, the present luminosity function can be identified with the
initial one). We used the luminosity function of Kroupa et al., which is
actually valid for population I stars, since uncertainties in the metal-poor
luminosity function are presently large although significant progresses have
been recently done (Kroupa \& Tout 1997). With this recipe, we find that
the systematic offset $X$\ is $X=0.17$~mag. However, $X$\ may be slightly
larger if the luminosity function for metal-poor stars is peaked at brighter
magnitudes than that for metal-rich objects (as indicated by both models and
observations: see Kroupa \& Tout 1997). A larger effect would be
due to a correlation between the masses of primaries and secondaries, as
suggested by Duquennoy \& Mayor (1991) for more massive binaries. If for
instance we had adopted a flat luminosity function for secondaries (very
likely a gross overestimate of the contribution by large mass secondaries), we
would have obtained $X=0.29$. 

The probability $p_i$\ that a star in our sample is a binary is not the same
for all stars in our sample: indeed, $p_i$\ is 1 for known binaries, close to 1
for suspected binaries (we will assume it is 1 also in this case), and quite
low for {\it bona fide} single stars. The probability $p_s$\ that a star is a
binary even if it has passed all the binarity tests is then $p_s=
(f-f_d)/(1-f_d)$, where $f$ is the real fraction of binaries and $f_d$\ is the
fraction of stars detected as binaries. As discussed in Section 2.4,
$f_d=0.41$\ for our sample. Kroupa et al. (1993) suggest that the fraction of
binary systems in the field is $f=0.6\pm 0.2$, implying $p_s=0.3\pm 0.3$. This
value is likely to be an overestimate in our case, because it does not take
into account that some of these binaries have a very faint companion (not
affecting the total magnitude and the colors); that a fraction of the binaries
are resolved so that the data of each component can be used (like e.g. for the
common proper motion pair HD134439-HD134440); and that there are {\it a priori}
selection effects against the presence of binaries in the sample (e.g. we could
not derive reliable metal abundances for tidally locked systems, which were
therefore dropped from the sample of stars used in the derivation of GC
distance moduli). 

\begin{figure} 
\centerline{\hbox{\psfig{figure=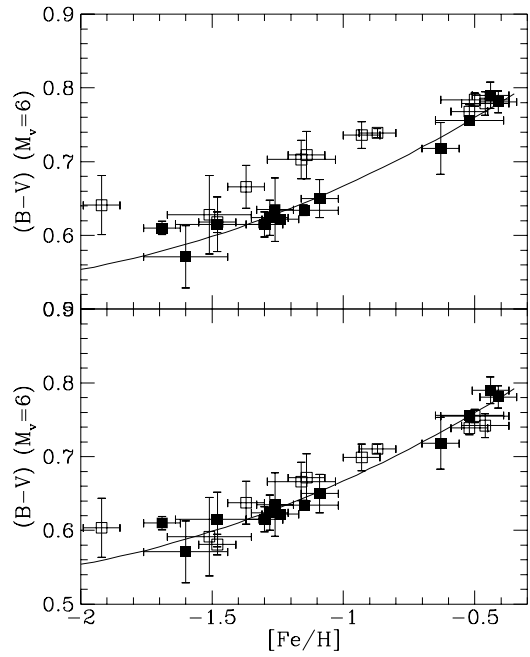,width=13.0cm,clip=}}}  
\medskip
\caption{The same as Figure~5, but with binaries included (open squares).
Individual binaries data are shown without (upper panel) and with the 
correction
for the contribution due to the secondary component (lower panel)} 
\label{f:fig3add} 
\end{figure}

On the whole, these arguments suggest values of $0.17\leq X\leq 0.29$\ and
$p_s=0.3\pm 0.3$. A more stringent constraint on the values
of $X$\ and $p_s$\ most appropriate for our subdwarf sample 
can be determined using two observational quantities: (i)
the systematic offset between the main sequence location derived from 
{\it bona fide} single stars and that obtained from 
known and suspected binaries; and (ii) the
observed scatter for individual {\it bona fide} single stars. In
Figure~\ref{f:fig3add} we show the location of individual binaries in the
[Fe/H]-$(B-V)(M_V=6)$\ diagram. Before any correction is applied (upper panel),
binaries are on average redder than {\it bona fide} single stars by $0.035\pm
0.007$~mag. This offset in color implies an offset of $0.176\pm 0.035$~mag in
magnitude. This is related to $X$\ and $p_s$ by the equation: 
\begin{equation}
p_s=1-(0.176\pm 0.035)/X
\end{equation}
On the other side, the observed small scatter of the 
{\it bona fide} single stars
around the mean relation of eq. (4) ($0.016\pm 0.005$~mag in $B-V$\ color) is
even smaller than the scatter expected from observational errors in the
parallaxes, colors and metallicities (making up an expected r.m.s of $0.027\pm
0.008$~mag). We may estimate that the scatter in color due to inclusion of
undetected binaries is $\leq 0.017$~mag at 95\% level of confidence and 
that with the same level of confidence ($\sim
2$~standard deviations) $0\leq p_s\leq 0.16$\ and $0.11\leq X\leq 0.21$. 
While these ranges of confidence are well consistent with those derived from
{\it a priori} arguments, they provide much severe constraints on 
$p_s$\ and $X$.

We thus estimate that {\bf systematic corrections of our distance moduli
for the possible presence of undetected binaries in our sample are $X~p_s=0.013
^{+0.020}_{-0.013}$~mag}, in the sense that our distances should be reduced by
this small amount. Distance moduli corrected for this effect are listed in
Column 8 of Table 3. 

Finally, we note that support to a small value of the systematic binary
correction to apply to our data is given by the following two arguments.
First, the good agreement between the position of single stars and binary 
systems in the [Fe/H]-$(B-V)_{(M_V=6)}$\ diagram of
Figure~\ref{f:fig3add} panel (b). Here a systematic correction of
$X=0.17$~mag to $M_V$ has been applied to individual binaries 
while no correction was applied to single stars. Second, the small 
difference between distance moduli derived from only the {\it bona fide} single
stars (Column 6 of Table 3), and those derived from the whole
sample of subdwarfs with $M_V>+5.5$\ after correcting each single star
according to its probability to be a binary (Column 10 of Table 3; the number
of stars used for each cluster is given in Column 9). We found that the
distances derived from the whole sample are on average smaller than those
derived using single stars alone by only $0.02\pm 0.01$~mag. 

On the basis of these tests, in the following discussion we will adopt the
distance moduli corrected for binarity (Column 8 of Table 3). It should be
recalled that the correction for binarity is a statistical one. Hence, the
error bars for the metal-poor clusters must be considered with caution, in
particular when few stars are available: uncertainties in the distances of the
most metal-poor clusters ([Fe/H]$<-1.8$) are clearly larger than represented by
the internal error. 

\subsection{Comparisons with other distance derivations}

\subsubsection{Reid's subdwarf fitting} 

The first comparison to be made is with the very recent distance determinations
using HIPPARCOS subdwarf parallaxes by Reid (1997). While the global procedure
and data source are similar, there are a number of differences (and we think
improvement) between our distance determinations and those by Reid. They are: a
larger subdwarf sample used here; a new analysis of the Lutz-Kelker
corrections; different magnitude ranges (here we limited ourselves to the
unevolved section of the MS); an improved analysis of the binary contamination;
new determination of subdwarf colors; revised cluster mean loci; different and
updated metal abundances (for both field stars and GCs); and different
reddening estimates for GCs. Following Reid's recommendation, we used his
$(m-M)_{12}$\ distances (true distance moduli) and mean reddening estimates in
the comparison. The mean residual (our estimates-Reid) is $-0.08\pm 0.04$~mag
(6 clusters, r.m.s. scatter of 0.09~mag), that is we found the clusters to be
closer to us than obtained by Reid (this difference corresponds to ages on
average older by 1~Gyr). However, our distance modulus for M5 is 0.04~mag
larger. Hence we expect a much smaller dependence of the horizontal branch
magnitude on metallicity than found by Reid (see below). 

\subsubsection{White dwarfs in NGC6752}

Renzini et al. (1996) used the white dwarf cooling sequence to derive a
distance modulus of $(m-M)_o=13.05\pm 0.12$\ for NGC6752. They assumed a
reddening value of $E(B-V)=0.04\pm 0.02$. Since the slope of the white dwarf
cooling sequence in the plane (F439W-F555W)-(F555W) is 5.9, the distance
modulus derived with our estimate of the reddening ($E(B-V)=0.035$) would be
$(m-M)_o=13.08\pm 0.12$, and hence $(m-M)_V=13.19\pm 0.08$. Our value is
$(m-M)_V=13.32\pm 0.08$. The two values agree within their error bar,
confirming the identification of the observed white dwarf cooling sequence with
the DA-one. 

\subsubsection{Absolute magnitude of the Horizontal Branch}

The absolute magnitude of the HB $M_V(HB)$\ has often been used to derive
distances and ages of globular clusters. There is however a rather hot debate
about the correct relation between $M_V(HB)$\ and the metallicity [Fe/H]. It is
then very interesting to derive this relation from our distance moduli which
are determined independently from the HB luminosity. However, there is some
ambiguity here because observations yield an average luminosity of the HB while
models usually give the absolute magnitude of the Zero Age HB (ZAHB). When
stars evolves off the ZAHB, they become brighter and change colors. Hence
$M_V(HB)$\ does not coincide with $M_V(ZAHB)$. Carney et al. (1992) give a mean
relation between $M_V(HB)$ (at the RR Lyrae color) and $M_V(ZAHB)$\ based on
observations of eight globular clusters. Models predict a smaller difference
between ZAHB level and the average magnitude of the RR Lyraes: Caloi, D'Antona
\& Mazzitelli (1997) quote a value of 0.06~mag, with a weak dependence on
metallicity. An intermediate result ($\sim 0.1$~mag) has been reached from the
analysis of synthetic HB models by Caputo \& Degl'Innocenti (1995). We will use
the following relation suggested by Sandage (1993): \footnote{Within this
discussion, we will use [Fe/H]+1.5 rather than simply [Fe/H] as the independent
variable; while this is just a matter of definition, it helps the reader to
distinguish between differences in the average absolute magnitude of the HB (a
matter related to the absolute ages of GCs) and differences in the slope (which
is more related to differential ages). The main aim of the present analysis is
to establish as well as possible the absolute ages of (the oldest) GCs; indeed
the error bar of the zero point of our M$_V$(ZAHB)-[Fe/H] relation is small. On
the other side, the use of a restricted number of clusters and the rather
strong dependence on the accuracy of the photometric calibrations and reddening
determinations, make the main sequence fitting technique not well suited to
derive the slope of this relation. This reflects in the rather large error bar
of the linear coefficient} 
\begin{equation}
M_V(HB)=M_V(ZAHB) - (0.05\pm 0.03) ({\rm [Fe/H]}+1.5) - (0.09\pm 0.04).
\end{equation}

$M_V(HB)$\ also does not coincide with the average magnitude of field RR Lyrae
of the same metallicity ($M_V(RR)$): this is because most GCs have few
variables (expected to be evolved of the ZAHB), while the field
variable population should be dominated by object still very close to ZAHB for
the metallicity range of GCs (see discussion in Carney et al. 1992). Finally,
the small slope with color of the HB should also be considered. 

The most recent collections of cluster $M_V(HB)$'s are those by Buonanno, Corsi
\& Fusi Pecci (1989), and Harris (1996). Harris $V(HB)$s\ are on average
brighter than Buonanno et al. $V(ZAHB)$\ by $0.03\pm 0.01$~mag, with no
distinct trend with metal abundance. Harris values refer to some average
magnitude of the HB, whose definition is not clearly stated in the catalogue.
Buonanno et al. identify their magnitudes with that of the ZAHB (as also
assumed in papers that make use of Buonanno et al. values : see e.g. Carney et
al. 1992, and Sandage 1993). This implies that Buonanno et al. should refer to
some lower envelope of the observed HBs. However, 
a close inspection of the original data considered by
Buonanno et al., as well as the use of $V(HB)$ estimates from recent 
papers in the literature, revealed that Buonanno et al. values should not be
intended as $V(ZAHB)$, but rather as $V(HB)$. The values of $V(ZAHB)$\ can be
derived from those of Buonanno et al. by applying eq. (6). 
   
Adopting for the errors associated to the involved quantities those listed in
Table~4 and a 0.06~dex uncertainty for the cluster metallicities 
(Column 3 of Table 3), the relation between
the absolute magnitude of the HB and metallicity for our nine clusters is: 
\begin{equation}
M_V(HB) = (0.17\pm 0.09)({\rm [Fe/H]}+1.5) + (0.40\pm 0.04)
\end{equation}

If we adopt for V$_{HB}$\ the values listed in Harris (1996) catalogue rather
than those from Buonanno et al., we get: 
\begin{equation}
M_V(HB) = (0.12\pm 0.09)({\rm [Fe/H]}+1.5) + (0.36\pm 0.04),
\end{equation}
where an uncertainty of 0.10 mag has been adopted for Harris $V_{HB}$\ values.
Finally, if we only consider the six clusters having most certain values of
$M_V(HB)$\ (that is if we eliminate the blue horizontal branch clusters NGC~288,
M13, and NGC~6752), we obtain:
\begin{equation}
M_V(HB) = (0.17\pm 0.09)({\rm [Fe/H]}+1.5) + (0.39\pm 0.04).
\end{equation}
where the $V(HB)$ values of Buonanno et al. have been used.
We see that while some uncertainty still exists in the
slope of the $M_V(HB)$--[Fe/H] relation 
(the uncertainty is even larger if the small number of metal-poor
comparison stars is taken into account), the zero point is well
determined. We will hereinafter adopt the relation obtained with Buonanno et
al. $V(HB)$'s (eq. 7). Combining Eq.s (7) and (6), we obtain the following
relation for $M_V(ZAHB)$: 
\begin{equation}
M_V(ZAHB) = (0.22\pm 0.09)({\rm [Fe/H]}+1.5) + (0.49\pm 0.04).
\end{equation}
Although most of the field RR Lyrae's should be quite close to the ZAHB, a small
average correction is required when considering the relation concerning these
objects. Following Caloi et al. (1997) we adopt an average correction of
0.06~mag (from the ZAHB) independent of metal abundance: 
\begin{equation}
M_V(RR) = (0.22\pm 0.09)({\rm [Fe/H]}+1.5) + (0.43\pm 0.04).
\end{equation}

The use of the above relationships has a direct impact on several astronomical
issues. For instance, since our value of $M_V$\ for field RR Lyraes at
[Fe/H]$=-1.9$ ($M_V=+0.34\pm 0.07$) is 0.10 mag brighter than the value quoted
by Walker (1992), we derive a distance modulus for the LMC of
$(m-M)=18.60\pm 0.07$\ (where the error is the statistical one at
[Fe/H]$=-1.9$). If this distance to the LMC is used (rather than the one
frequently adopted from Cepheids: $(m-M)=18.50\pm 0.10$), the extragalactic
distance scale increases (and estimates of the Hubble constant
decrease) by 5\% (for instance, the value of $H_o$\ derived from SN~Ia by Hamuy
et al. 1996, would change from $63.1\pm 3.4\pm 2.9$\ to $59.9\pm 3.2\pm 2.8$).
Further discussion on the LMC and M31 distance moduli after HIPPARCOS can be
found in Feast and Catchpole (1997). We only remark that the distance modulus
for the LMC based on the HIPPARCOS calibrations is $(m-M)=18.70\pm 0.10$\ from
Cepheids (Feast \& Catchpole 1997), and $(m-M)=18.54\pm 0.2$\ from Miras (van
Leeuwen et al. 1997) in excellent agreement with our determination. The most
recent determination from the expanding ring around the SN 1987a is
$(m-M)=18.58\pm 0.03$\ (Panagia et al. 1997). 

To properly compare our derived $M_V$(HB)--[Fe/H] relationship with previous
results, one should separate the discussion of the slope from that of the
zero-point. Though we postpone a complete analysis of this specific issue to a
future paper, we note that the slope of our relation, albeit quite uncertain,
is somewhat smaller than (but still compatible with) that found by Sandage
(1993), usually taken as the proto-type of the so-called "steep" slopes: 
\begin{equation}
M_V(RR) = 0.30 ({\rm [Fe/H]}+1.5) + 0.49;
\end{equation}
reasons to prefer a lower value of the slope (similar to ours) are given by
Carney et al. (1992; see however Sandage 1993 for a different viewpoint).
Typical examples of "shallow" slopes are those provided by the 
BW analysis of field
RR Lyraes, like for instance: 
\begin{equation}
M_V(RR) = (0.16\pm 0.03) ({\rm [Fe/H]}+1.5) + (0.78\pm 0.03)
\end{equation}
and:
\begin{equation}
M_V(RR) = (0.19\pm 0.03) ({\rm [Fe/H]}+1.5) + (0.68\pm 0.04)
\end{equation}
from Jones et al. (1992) and Clementini et al. (1995), respectively. Finally,
a slope similar to the value presented here has been obtained from the HBs
of the globular clusters observed in M31 with HST (Ajhar et al. 1996, Fusi
Pecci et al. 1996): 
\begin{equation}
M_V(HB) = (0.13\pm 0.07) ({\rm [Fe/H]}+1.5) + (0.75\pm 0.09),
\end{equation}
although the available sample is still too poor in both cases to firmly
establish the dependence. 

When discussing the zero point of the $M_V(HB)$--[Fe/H] relation, we recall
that the values determined from the pulsational properties of RR Lyrae are
still somewhat uncertain (see e.g. Carney et al. 1992). The value obtained by
Sandage (1993, to be compared with that of $M_V(RR)$, see Eq. 11) is similar to
ours (the difference being only 0.06~mag). As to the M31 clusters, it should be
reminded that the zero point of Eq. (15) is based on a distance modulus of
$(m-M)_0=24.43$\ for M31 (Freedman \& Madore 1990). If the value of
$(m-M)_0=24.77$\ suggested by Feast \& Catchpole (1997) and consistent with the
HIPPARCOS distance to the galactic Cepheids is adopted, the zero point would
become $0.41\pm 0.09$; and it would be $0.51\pm 0.09$\ if the extragalactic
distance scale deduced here is adopted. These values (to be compared with
$M_V(HB)$, see Eq. 7) are in good agreement with that determined in the present
paper. 

On the other side, the present zero-point is 0.2 to 0.3~mag brighter
than any figure obtained so far from the application of the BW-method to field
RR Lyraes, as reviewed and discussed for instance by Carney et al. (1992). The
direct comparison between the BW absolute magnitudes determined for cluster RR
Lyraes and our HB magnitudes yields a similar difference. In fact, we find
$M_V(HB)=0.25$\ and 0.51 for M92 and M5, respectively, against values of 0.45
and 0.66 determined using the BW-method (Storm et al. 1994). On average,
the $M_V(HB)$\ values determined from the subdwarf fitting is 0.18~mag brighter
than that derived using the BW-method. 

It may also be worth recalling that the zero-point obtained via the BW-method
agrees quite nicely with the absolute magnitudes for field RR Lyraes determined
using the statistical parallaxes ($M_V(RR)=+0.71\pm 0.12$\ for [Fe/H]=$-1.6$
and  $M_V(RR)=+0.79\pm 0.30$ for [Fe/H]=$-0.76$: Layden et al. 1996). It is
thus quite evident that a deeper insight into these procedures and analyses is
urged to explain the reason of the current discrepancy which has, by itself, a
huge impact on the GC ages (a luminosity difference of $\sim 0.07$~mag
corresponds to about 1 Gyr !). 

\begin{figure} 
\centerline{\hbox{\psfig{figure=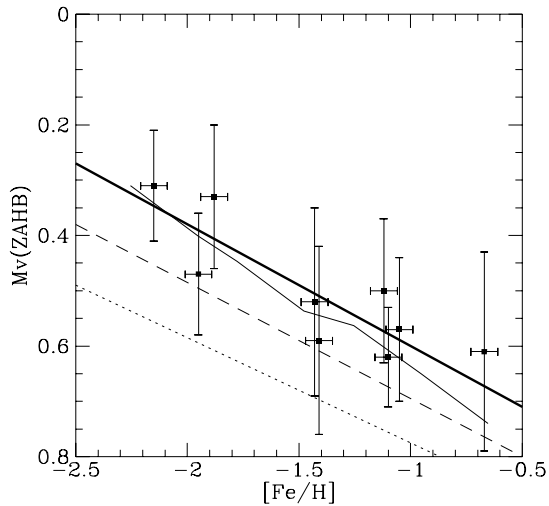,width=13.0cm,clip=}}}  
\medskip
\caption{Runs of the $M_V$(ZAHB) against [Fe/H] for the programme cluster using
our distance moduli and $V(ZAHB)$\ from Buonanno et al. (1989), corrected for
the difference between $M_V$(ZAHB) and $M_V$(HB) (see text). The solid thick
line is the weighted least square fit line through the points. For comparison,
we also show the predictions based on the horizontal branch models by Caloi et
al. (1997: solid thin line), VandenBerg (1997: dotted line), and Salaris et al.
(1997: dashed line)} 
\label{f:fig6} 
\end{figure}

Turning to the HB theoretical models, the available scenario has been quite
homogeneous until recently. In fact, most computations were leading to similar
slopes ($\sim 0.15-0.20$) and zero-points ($\sim 0.8$, using our definition).
The reason for this similarity being essentially the use of the same input
physics (leading to almost identical core masses for the red giant branch and,
in turn, HB stars), and similar transformations to the observational plane.
The luminosity of the HB models is raised by about 0.1 mag by using updated 
input physics
(opacities and equation of state). Typical results are those very recently
obtained by VandenBerg (1997): 
\begin{equation}
M_V(ZAHB) = 0.19 ({\rm [Fe/H]}+1.5) + 0.68,
\end{equation}
and Salaris et al. (1997):
\begin{equation}
M_V(ZAHB) = 0.21 ({\rm [Fe/H]}+1.5) + 0.59.
\end{equation}
While predicting very similar slopes, these models give zero points 
values that are respectively 0.19 and 0.10~mag fainter than that derived from the main sequence fitting.
However, a couple of years ago Mazzitelli et al. (1995) proposed that the core
masses are about 0.01 M$_\odot$\ larger then previously assumed for HB stars.
This suggestion has been further supported by a new paper by the same group
(Caloi et al. 1997), and very recently by other investigators
too (Straniero and Chieffi 1997). A larger core mass implies brighter
luminosities; the latest HB models computed by Caloi et al. (1997) yield: 
\begin{equation}
M_V(ZAHB) = 0.26 ({\rm [Fe/H]}+1.5) + (0.515\pm 0.07),
\end{equation}
where the error bar refers to the transformations used (VandenBerg 1992 giving
brighter magnitudes than Kurucz 1993), and the slope represents just an average
slope as their $M_V$(HB)--[Fe/H] relation is not linear. Both the slope and the
constant term of the Caloi et al. relation agree well with our eq. (11) (see
also Figure~\ref{f:fig6}). 

\subsubsection{Astrometric distances to GCs}

Distances to globular clusters may be derived by comparing the distribution of
the radial velocities of the clusters with that of the internal proper motions,
using King-Michie type dynamical models. While results for individual clusters
derived by this procedure are affected by large error bars, and depend on
cluster dynamical models, they are totally independent from stellar evolution
models, and provide then an useful comparison. Rees (1996) gives updated
distances based on this technique for ten GCs: four of them are in our list. If
results for these four clusters are weighted according to their error bars, the
astrometric distances are on average larger than those derived from subdwarf
fitting by $0.01\pm 0.12$~mag. However, if the astrometric distance to 47~Tuc
(which is suspect because rotation, neglected in the analysis, may be
important) is dropped, the astrometric distances would be shorter than those
derived from subdwarf fitting by $0.19\pm 0.17$~mag. Besides, Rees quotes a
value of $M_V(HB)=0.61\pm 0.05$\ at [Fe/H]=$-1.6$. We redetermined the value at
[Fe/H]$=-1.5$, using our slope for the $M_V(ZAHB)-$[Fe/H] relation and our
metallicity scale, and we found $M_V(ZAHB)=0.62\pm 0.10$\ (the original small
error bar quoted by Rees seems to be underestimated, since it does not agree
with the errors for individual clusters listed in his Table~1); however, the
value is lowered to $M_V(ZAHB)=0.50\pm 0.09$\ if 47~Tuc is included. While
astrometric distances to GCs might be somewhat shorter than those derived from
subdwarf fittings, more precise values are required to settle this point. 

\section{GLOBULAR CLUSTER AGES}

Ages for the programme clusters were derived from the absolute magnitude of the
turn-off, $M_V(TO)$. We used the distance moduli of Table~3 and the $TO$\
apparent magnitudes, $V(TO)$\, listed by Buonanno et al. (1989).
\footnote{Adoption of $V(TO)$\ luminosities from Buonanno et al. rather than
those directly derived from the mean-loci used for distance derivations
introduces some inconsistency. However, we regard this effect as minor with
respect to the rather large uncertainties related to the estimate of $V(TO)$\
from a given set of mean loci. Anyway, we note that very similar results are
obtained using the turn-off magnitudes by Chaboyer, Sarajedini \& Demarque
(1992).} 

$M_V(TO)$ is an age indicator much less dependent on details of
the models used than those based on colors, which are heavily affected by the
assumptions about opacities, the treatment of convection, the equation of
state, and the color transformations (see Chaboyer
1995, and VandenBerg et al. 1996 and references therein, for a 
thorough discussion of these topics). For the same reason
it should be preferred to the direct fitting of the isochrones and the observed
main sequences (Renzini 1991). 

On the other side, the determination of the exact location of $M_V(TO)$\ is made
difficult by its own definition (the bluest point of the main sequence, and
hence the region where mean loci are vertical). Typical errors quoted for
$V(TO)$\ are $0.05-0.10$. If we combine quadratically this error to those
in the distance moduli (from 0.07 to 0.09~mag: see Table~3), we find that
errors in $M_V(TO)$\ are $0.09 - 0.13$~mag, at least. We will hereinafter
assume a typical error bar of $\pm 0.11$~mag, which implies random errors of
12\% in ages. 

Table~5 lists the main cluster parameters, as well as the ages we derived from
different sets of isochrones. In the isochrones labeled as MLT, convection has
been modeled using the Mixing Length Theory, while those labeled as CM use the
Canuto-Mazzitelli (1991) theory. On any other respect, these isochrone sets are
quite similar to each other: they use updated equation of state (including
Debye screening), opacities from the Los Alamos group, and color
transformations according to Kurucz (1993). 

Inspection of Table~5 shows that differences between ages derived from
different isochrone sets are rather large, even if we restrict to models that
use MLT-convection. 

The average difference between isochrones providing the oldest ages (D'Antona
et al. 1997), and those giving the youngest ages (Bertelli et al. 1997) is
$\sim 1.7$~Gyr. A large fraction of this spread is simply due to different
assumptions about the solar absolute magnitude $M_V$. 
\footnote{ Some confusion may arise here because it is often said that these
differences are due to the definition of the zero-point of bolometric
magnitudes $M_{\rm bol}$\ or of the bolometric corrections. If this were
true, derived ages would be arbitrary, because the zero-points of both
bolometric magnitudes and bolometric corrections are arbitrary. Luckily enough,
this is not the case as bolometric magnitudes and bolometric corrections are 
only used as intermediate steps in the transformations, and the arbitrariness 
in their zero-points cancels out when consistent values are used.} 

To better understand this point, we recall that when computing isochrones, the
solar model is calibrated against the observed luminosity by adjusting the
solar helium content, a quantity that is only loosely constrained by
observations. However, the helium content used for metal-poor isochrones is
actually independent of the solar value, and it is rather derived on
nucleosynthesis grounds (cosmological and sometimes galactic), supported by
observations of extragalactic metal-poor HII regions (see e.g. Izotov et al.
1997) or by the application of the so-called "R-method" (Iben 1968,
Buzzoni et al. 1983). Therefore, the solar luminosity calibration is not 
relevant for the ages of globular clusters: {\it i.e.} no parameter is 
adjusted when deriving the isochrone turn-off luminosities.

On the other hand, globular cluster observations yield magnitudes and not
luminosities, and some transformation from the theoretical $\log {L/L_\odot}$\
to the observational $M_V$\ quantity must be adopted. The zero-point of this
transformation is indeed based on the Sun, since this is the star for which
both quantities can be derived with the highest accuracy from observations. 

All isochrone sets assume the same value for the solar luminosity of
$L_\odot=3.86\,10^{33}$~erg~s$^{-1}$, but there are differences in the assumed
value for $M_{V_\odot}$, with values ranging from 4.78 to 4.85. By itself, this
implies a spread of $\sim 1$~Gyr in the derived ages, isochrones computed with
a brighter value for $M_{V_\odot}$\ yielding older ages. 

However, the value of $M_{V_\odot}$\ is not arbitrary, being constrained by
observations. The best value for this quantity is $M_{V_\odot}=4.82\pm 0.02$\
(Hayes 1985). The error bar translates into an uncertainty of only $\pm
0.3$~Gyr in the ages, if all isochrones were corrected to the same value of
$M_{V_\odot}$\ as determined by Hayes (1985). This can be done {\it a
posteriori} because it is simply a different definition of the zero-point of
the $\log {L/L_\odot}-M_V$\ transformation, and it has no other effect on the
models. In the following discussion, we will apply this correction to the ages
listed in Table~5. 

After this correction has been included, the spread in the ages is reduced to
$\sim 0.8$~Gyr. This residual spread is due to different assumptions about the
equation of state, opacities, helium and metal contents, etc. While a thorough
discussion of these assumptions might help to further reduce the spread, it is
beyond our expertise and the scopes of the present paper; we will thus keep the
spread of $\pm 3$\% around the mean value as one of the possible systematic
errors in our ages. 

Turning to the cosmological purposes, what matters is actually the age of the
{\it oldest} globular clusters. This may well be different from the mean age of
globular clusters. However as mentioned above, random errors are not
negligible, and they may be even larger than the quoted error bar of $\pm 12$\%
if allowance is given for possible calibration errors in the photometry. To
this respect we note that whether the original
Penny \& Dickens (1986) photometry for NGC6752 had been used, the age for 
this cluster would
have been increased by 1.6 Gyr. It seems thus wiser to average results for
different clusters, rather than to rely on a single object. Of course, there is
some arbitrariness in the selection of the clusters used to estimate this
average. Our procedure was as follows. 

The last column of Table~5 gives a rough classification of the cluster HB
morphology, clusters are divided into four groups: Oosterhoff I (OoI),
Oosterhoff II (OoII), Blue Horizontal Branch clusters (BHB), and Red Horizontal
Branch clusters (RHB). Oosterhoff I and II clusters have RR Lyraes with
different mean periods ($\sim 0.55$~d and $\sim 0.65$~d respectively:
Oosterhoff 1944). 

From our data, we found some scatter for the ages of Oosterhoff II clusters;
however this scatter is not significantly larger than the expected error bar.
Furthermore, the close similarity of the c-m diagrams for the metal-poor
clusters strongly support a common age (see the thorough discussion by Stetson,
VandenBerg \& Bolte, 1996, and panel {\i b} of Figure~\ref{f:fig5}). In the
following discussion we will then assume that Oosterhoff II clusters are indeed
coeval, and will attribute the scatter to observational errors. A similar
argument can be made for the BHB clusters (see panel c of Figure~\ref{f:fig5}).

On the other side, taken at face value, our age estimates for the Oosterhoff I
clusters in the sample (M5 and NGC362) are $\sim 2.4\pm 1.2$~Gyr lower than
those for the blue horizontal branch clusters (BHB: M13, NGC288, NGC6752),
which in turn are similar to those for the Oosterhoff II clusters (M92, M68,
M30). A rather low age is found also for 47~Tuc. The reality of this
age-difference is argument of hot debate. As well known, the Yale group (Zinn,
1993; Lee, Demarque \& Zinn, 1994; Chaboyer et al. 1992) strongly favours the
existence of an age difference, that would explain the different
HB-morphologies (but not important details of the color distributions: see
e.g. Fusi Pecci et al. 1993, Buonanno et al. 1997). On the other side, other
groups (see e.g. Stetson et al. 1996 and references therein) question even the
reality of the large difference in age put forward by the Yale group for the
pair NGC288-NGC362 (see e.g. Catelan \& de Freitas Pacheco 1994). It should
be noted here that our age indicator, selected to be a robust indicator of {\it
absolute} ages, is not well suited for the derivation of {\it relative} ages,
being strongly dependent on the assumed reddening and on the accuracy of the
photometry for each cluster. Given the small numbers, the statistical error
bars are not very significant. We remark that the very low ages we found for
the Oosterhoff I clusters is heavily weighted by the extreme result for NGC~362
($t\sim 8.6$~Gyr); however, our distance modulus for this cluster may be too
large, as suggested by the bright magnitude of $M_V(HB)=0.39\pm 0.13$\
we get for the HB. The value given by our mean relation (Eq. (7)) would be
$M_V(HB)=0.46$\, implying an age $\sim 1.0$~Gyr larger, much closer to the
average value for the remaining clusters\footnote{This discrepant result could
be explained if the mean loci used for NGC362 are still too blue by 0.01~mag,
either due to errors in the photometry or in the assumed reddening}. Hence,
although our data seems to give some support to the age difference, we think
that more data with higher accuracy and for a larger number of clusters 
are necessary to properly
address this problem. Anyway, it seems wiser to exclude the Oosterhoff I
clusters and 47~Tuc from our estimate of the age of the oldest globular
clusters in our sample. We thus identify the group of {\it bona fide} old
clusters with the Oosterhoff II and the BHB clusters. 

Average ages for our {\it oldest} globular clusters obtained with different
isochrone sets are given in Table~6. Here we listed for each isochrone set
the value of $M_{V\odot}$\ originally assumed, the mean ages we derived from the
isochrone sets as they are tabulated by the authors, and the mean ages computed
after correction to Hayes (1985) $M_{V\odot}$\ value. The r.m.s. scatters of
ages for individual clusters (from 1.0 to 1.2~Gyr depending on the isochrone
set, that is from 8 to $10$\%) is even smaller than the expected error bar.
Although we cannot exclude the possibility that some clusters are older than
the average, this is not required by our observations. We will then assume that
these average values are the best guess for the age of the oldest globular
clusters. 

If we use isochrones based on the MLT-convection, the mean age for the six
{\it bona fide} old clusters is $11.8\pm 0.6 \pm 0.4$~Gyr, where the first
error bar is the standard deviation of the mean values obtained for different
clusters, and the second error bar is the spread of ages derived from different
isochrone sets. However, as discussed by Chaboyer (1995), VandenBerg et al.
(1996), and Chaboyer et al. (1996), the non-formal error bar is surely larger
than these values. On the other side, a simple combination of the error bars is
not very meaningful, because we are combining statistical errors (standard
deviations) with maximum errors (as given by models), and because it is
unlikely that all errors combine in the same direction. This point is discussed
at length in Chaboyer et al. (1996). To better quantify the error bars, we
adopted a procedure very similar to that considered by these authors:
essentially, it is assumed that a suitable statistical distribution of errors
can be used for each source of uncertainty. A MonteCarlo procedure is then used
to derive the distribution of total errors and to provide the statistical
interval of confidence (95\% range). 

The following sources of uncertainty were considered:

\begin{itemize}
\item {\bf Internal errors:} the internal errors may be estimated from the
scatter of values for individual clusters; we found 0.4~Gyr. We assumed a
gaussian distribution with this standard deviation. 

\item {\bf Solar $M_V$:} The age error due to uncertainties in the assumed
value of $M_{V_\odot}$\ is 0.3~Gyr; we assume a gaussian distribution with this
standard deviation. 

\item {\bf Binaries:} Uncertainties in the statistical correction for undetected
binaries is given in Section 5.1 ($^{+0.02}_{-0.01}$~mag). This translates into 
an error bar of $^{+0.3}_{-0.2}$~Gyr on the cluster ages. We assumed a gaussian
distribution, truncated at $-$0.2 (the correction cannot be negative).

\item {\bf Reddening scale:} As mentioned in Section 4, an error of $\pm
0.01$~mag in the zero-point of the reddening scale yields an age uncertainty of
$\sim 1$~Gyr in the mean ages. Again we assumed a gaussian distribution with
this value for the standard deviation. 

\item {\bf Metal abundance scale:} The use of the new metallicity scale by
Carretta \& Gratton (1997) in place of Zinn and West's one coupled with the
inclusion of the appropriate $\alpha-$enhancements has a strong impact on the
derived ages. Leaving aside consistency between cluster and field star
abundances (which might affect distance derivations), the whole abundance scale
is uncertain by some 0.1~dex, which implies age uncertainties of about 0.4~Gyr.
We assumed a uniform distribution with these extreme values.

\item {\bf Stellar model code:} Our age estimates is an average of the values
given by the various sets of theoretical isochrones. We assumed a uniform
distribution over the range $-0.4$\ to 0.4~Gyr. 

\item {\bf Convection description:} D'Antona et al. (1997) computed models
using both the traditional MLT and the CM theory of convection. The
Canuto-Mazzitelli (1991) theory accounts for the full spectrum of eddy sizes,
with respect to the single value adopted in the MLT. However, both theories
have free parameters that must be calibrated against observations for the Sun.
Keeping everything else constant, the mean age for the oldest globular clusters
derived using the CM-convection is about 0.4 Gyr smaller than that derived using
the MLT-models. We assumed a uniform distribution with extremes $-0.4$~Gyr and 
0.

\item {\bf He-sedimentation:} Ages would be further reduced by $\sim 1$~Gyr if
allowance is given for helium sedimentation during the main sequence lifetime
(Chaboyer \& Kim 1995; D'Antona et al. 1997). This is to be interpreted as a
maximum error, since larger effects are difficult to be reconciled with the
so-called Li-plateau (for a discussion, see VandenBerg et al. 1996).
He-sedimentation would also help reproducing the color difference between the
turn-off and the subgiant branch, which would else be too large in standard
models with such small ages and high metallicities (see e.g. Brocato,
Castellani \& Piersimoni 1997). We assumed a uniform distribution between
$-1$~Gyr and 0. 

\end{itemize}

We then run 1,000 Monte Carlo extractions, using random generated
values for each of the above mentioned sources of uncertainty. In this way, we 
find that the 95\% probability range of confidence is from 9.3 to 13.9~Gyr.

In summary, we find that the age of the oldest globular clusters is:
$${\rm Age}={11.8^{+2.1}_{-2.5}}{\rm Gyr}.$$ 

\section{SUMMARY AND CONCLUSIONS}

Very accurate distances and chemical abundances have been determined for a
sample of about 150 dwarfs with $-2.5<$[Fe/H]$<0.2$, using data from the
HIPPARCOS satellite and high S/N, high dispersion spectra. On the whole, we can
then exploit a rather large homogeneous sample with standard errors of $\Delta
M_V\sim 0.1$~mag and $\Delta $[Fe/H]$\sim 0.07$~dex. Oxygen and
$\alpha-$element (Mg, Si, Ca, and Ti) abundances were also determined for most
of the stars in the sample. We found that all stars with [Fe/H]$<-0.5$\ have
overabundances of [O/Fe]=$0.38\pm 0.13$\ and [$\alpha$/Fe]=$0.26\pm 0.08$\
respectively. 

Various sources of systematic errors were examined: in particular, we estimated
the relevance of Lutz-Kelker corrections and of undetected binaries in the
sample. Appropriate corrections were included in our results. 

The subdwarf sample allowed us to determine the {\it absolute} location of the
main sequence as a function of metallicity, using a procedure which depends
only in part on theoretical models. These semi-empirical main sequences have
then been compared with the best available photometric data for nine carefully
selected, low-reddening Galactic globular clusters. 

The selected clusters have abundances determined from high resolution spectra
of individual giants (Carretta \& Gratton 1997), following exactly the same
procedure used for the field subdwarfs. These high dispersion analyses indicate
that cluster stars have O and $\alpha$-element excesses similar to those found
for the field stars. Reddenings for four of the nine clusters were redetermined
using Str\"omgren photometry of early type stars projected toward the same
direction of the clusters. Once averaged with previous reddening estimates, our
values of $E(B-V)$\ have typical errors of 0.008~mag. 

With these recipes, distance moduli were derived for the clusters with typical
errors of 0.08~mag. Ages were then deducted using the absolute magnitude of the
turn-off point, a theoretically robust indicator, which depends only marginally
on details of the models used. The main conclusions are: 

\begin{itemize}

\item The relation between the HB absolute magnitude and metallicity for
the nine programme clusters is:
$$M_V(ZAHB) = (0.22\pm 0.08)({\rm [Fe/H]}+1.5) + (0.49\pm 0.04)$$
This relation is consistent with some of the most recent theoretical models
(Caloi et al 1997). The slope of the relation agrees with that determined from
the Baade-Wesselink method (see e.g. the discussion in Carney et al. 1992), and
is less steep than that given by Sandage (1993). The constant term agrees fairly
well with that given by Sandage, and it is 0.2 to 0.3~mag brighter then that
obtained by the Baade-Wesselink method. Our best estimate for the distance
modulus of the LMC is $(m-M)_0=18.60\pm 0.07$, in good agreement with the
recent values of $18.70\pm 0.10$\ and $18.54\pm 0.2$\ derived by Feast and
Catchpole (1997) and van Leeuwen et al. (1997), respectively, from HIPPARCOS
parallaxes of Galactic Cepheid and Mira variables. 

\item The age of the oldest globular clusters (Oosterhoff II and BHB) is:
$${\rm Age} = 11.8^{+2.1}_{-2.5}{\rm Gyr}$$
(the error bar corresponds to the 95\% range of confidence). The r.m.s. scatter
of individual values (from 1.0 to 1.2 Gyr, i.e. from 8 to 10\%, depending on
the adopted isochrone set) agrees with expectations of the observational errors
in the distance moduli and in the estimate of the magnitude of the turn-off
point (i.e. we did not find compelling evidence for an age spread amongst this
subset of clusters). 

We found that the Oosterhoff I clusters in our sample (M5, NGC362) have ages
$\sim 2.4\pm 1.2$~Gyr younger than the blue horizontal branch clusters (BHB:
M13, NGC288, NGC6752), which in turn have ages similar to the Oosterhoff II
clusters (M92, M68, and M30). A rather young age is found also for the red
horizontal branch cluster 47 Tuc. While this age difference might explain, at
least partially, the different mean color of the observed horizontal branches,
it may well be due to observational errors in the photometries and in the reddening
values used. Hence, identification of age as {\it the} second parameter
(currently an argument of hot debate amongst investigators) should wait for
results based on age indicators better suited to this purpose. However, it
seemed wiser to exclude the Oosterhoff I clusters from the estimate of the age
of the oldest globular clusters. 
\end{itemize} 

\bigskip
$\bigoplus$ {\it  Why do cluster ages become younger?}

To have a quick insight into this big issue one has to recall that : 
an increase of the TO-luminosity by $\sim 0.07$~mag {\it or}  of the
TO-color by $\sim 0.01$ mag roughly yields a corresponding increase of the 
age by about
1 Gyr. Moreover, the adoption of different assumptions in the transformations
from the theoretical to the observational plane and in the model computation
(as far as treatment of convection, sedimentation, etc. are concerned) may have
a significant impact on the cluster-age game. 

We get ages much {\it younger} than classically obtained mainly because of: 

\begin{itemize}

\item An increase by $\sim 0.2$~mag in the distance moduli, due to the use of
the new HIPPARCOS parallaxes, which yields a decrease of $\sim 2.8$~Gyr over
the whole metallicity range covered by the programme clusters. 

\item A variation in the adopted metallicity scale (from Zinn and West 1984 to
Carretta and Gratton 1997), which produces distance moduli larger by $\sim
0.15$~mag, and, in turn, ages younger by $\sim 2$~Gyr for the
intermediate-metallicity clusters. Note that almost no age difference is
implied by this new metallicity-scale at the very metal-poor and metal-rich
extremes. 

\item A full consistency in the metal abundance determination of cluster and
field stars, including the $\alpha-$elements, which leads then to compute ages
based on the {\it global} metallicity, and not just on [Fe/H]. This yields
smaller ages, by about 1-2 Gyr (somewhat depending on metallicity) than
age determinations done before 1990 (the effect of an enhancement of the
$\alpha-$elements was generally considered in recent age determinations for
GCs: see e.g. Carney et al. 1992, VandenBerg et al. 1996). 

\end{itemize}
Besides, one could further decrease ages by about 1--1.5 Gyr considering MLT
vs. CM convection and He-sedimentation. 

\bigskip
$\bigoplus$ {\it Final conclusion: impact on cosmology}

To translate the present age estimate for GCs into a lower limit to the age of
the Universe, the epoch of cluster formation should be known. Unfortunately,
this is not the case. Loeb (1997) recently suggested that the first star
clusters formed at $z=10$\ or even $z=20$. This can correspond to a delay of
less than 0.2 Gyr. However, we assumed more conservatively a minimum delay of
0.5 Gyrs from the birth of the Universe before the formation of globular
clusters. Hence our age estimate is compatible with an Einstein-de Sitter model
if $H_0\leq 64$ km s$^{-1}$Mpc$^{-1}$, and $H_0\leq 83$ km s$^{-1}$Mpc$^{-1}$\
in a flat Universe with $\Omega_m=0.2$. Within the framework of inflationary
models (even in the restricted but more elegant solution of the Einstein-de
Sitter universe), the presently determined age for the globular clusters is
then consistent with current estimates of the Hubble constant, even without the
$\sim 5$\% reduction which is given by the adoption of the present distance
scale, or that proposed by Feast \& Catchpole 1997. We conclude that {\bf at
the present level of accuracy of globular cluster ages, there is no discrepancy
with standard inflationary models for the Universe}. 

On the other hand, our best value for the age of globular clusters (plus 0.5
Gyr for cluster formation) yields a best estimate of 12.3~Gyr for the age of
the Universe: this requires a Hubble constant of 51.7~km~s$^{-1}$Mpc$^{-1}$ for
an Einstein-de Sitter universe, and of 66.4~km s$^{-1}$Mpc$^{-1}$\ for a flat
Universe with $\Omega_m=0.2$. For comparison, if the present distance scale is
adopted for extragalactic observations (that is 5\% larger than that usually
adopted), the value of $H_o$\ derived from SN~Ia (Hamuy et al. 1996) would
become $59.9\pm 3.2\pm 2.8$~km~s$^{-1}$Mpc$^{-1}$, and that from Cepheids in
the Virgo cluster (Freedman et al. 1997) would be $H_0=69\pm
10$~km~s$^{-1}$Mpc$^{-1}$. 

\acknowledgments
{The Hipparcos data used in the original sample were the result of the FAST
proposal n. 022; we are grateful to P.L. Bernacca for allowing us to have early
access to them and for continuous help in the related procedures. We wish to
thank P. Bertelli, A. Bressan, F. D'Antona, F. Lucchin and S. Ortolani for many
useful discussions and for having provided data in advance of publication; D.A.
VandenBerg  and S. Chieffi for sending their new tracks; C. Sneden and G.
Cutispoto for their collaboration in the data acquisition; I.N. Reid for having
provided an electronic version of his paper in advance of publication; F. Pont
for stimulating discussions about the problem of binary contamination, at the
ESA Hipparcos meeting in Venice. We are indebted to the anonymous referee for
many useful suggestions, that contribute to improve the paper. E.Carretta
gratefully acknowledges the support by the Consiglio Nazionale delle Ricerche.
The financial support of the {\it Agenzia Spaziale Italiana} (ASI) is also
gratefully acknowledged. This research has made use of the SIMBAD data base,
operated at CDS, Strasbourg, France.}

\scriptsize

\begin{table}
\caption{Basic data for the field stars}
\begin{tabular}{rrrlccccccl}
\tableline
\tableline
 HIP.&HD/&$\pi$&$\delta \pi/\pi$&$V_0$&$M_v$&$(B-V)_0$& [Fe/H] &
[O/Fe] & [$\alpha$/Fe] & Note \\
 No.& Gliese&(mas)\\
\tableline
\\
\multicolumn{11}{c}{Stars in the original sample}\\
\\
   999&G030-52& 24.69&0.049&$8.515\pm 0.075$&$5.48\pm 0.10$&
$0.766\pm 0.024$&$-0.56\pm 0.13$&      &      & SO \\
 14594&  19445& 25.85&0.044&$8.050\pm 0.010$&$5.11\pm 0.09$&
$0.454\pm 0.018$&$-1.91\pm 0.07$& 0.56 & 0.38 &    \\
 15797&G078-33& 39.10&0.032&$8.971\pm 0.009$&$6.93\pm 0.07$&
$0.982\pm 0.002$&$-0.41\pm 0.07$&      & 0.16 &    \\
 16404&G246-38& 17.58&0.087&$9.910\pm 0.000$&$6.14\pm 0.18$&
$0.660\pm 0.010$&$-1.92\pm 0.07$&      &      & AB \\
 38541&  64090& 35.29&0.029&$8.280\pm 0.010$&$6.02\pm 0.06$&
$0.620\pm 0.004$&$-1.48\pm 0.07$& 0.46 & 0.32 & SB \\
 38625&  64606& 52.01&0.036&$7.430\pm 0.000$&$6.01\pm 0.08$&
$0.736\pm 0.006$&$-0.93\pm 0.07$& 0.78 & 0.22 & SB \\
 57939& 103095&109.21&0.007&$6.425\pm 0.005$&$6.61\pm 0.02$&
$0.754\pm 0.004$&$-1.24\pm 0.07$& 0.37 & 0.23 &    \\
 60956& 108754& 19.20&0.059&$9.009\pm 0.019$&$5.41\pm 0.12$&
$0.704\pm 0.003$&$-0.58\pm 0.07$& 0.42 & 0.14 & SB \\
 62607& 111515& 30.12&0.030&$8.139\pm 0.008$&$5.52\pm 0.06$&
$0.687\pm 0.003$&$-0.52\pm 0.07$& 0.29 & 0.10 & S? \\
 66509& 118659& 18.98&0.064&$8.820\pm 0.010$&$5.20\pm 0.14$&
$0.674\pm 0.002$&$-0.55\pm 0.07$& 0.51 & 0.08 &    \\
 72998& 131653& 20.29&0.074&$9.512\pm 0.002$&$6.05\pm 0.16$&
$0.720\pm 0.000$&$-0.63\pm 0.07$& 0.36 & 0.31 &    \\
 74234& 134440& 33.68&0.050&$9.441\pm 0.001$&$7.08\pm 0.11$&
$0.853\pm 0.000$&$-1.28\pm 0.07$&      & 0.15 &    \\
 74235& 134439& 34.14&0.040&$9.073\pm 0.002$&$6.74\pm 0.08$&
$0.773\pm 0.000$&$-1.30\pm 0.07$&      & 0.29 &    \\
 78775& 144579& 69.61&0.008&$6.660\pm 0.000$&$5.87\pm 0.02$&
$0.734\pm 0.004$&$-0.52\pm 0.13$&      &      &    \\
 81170&149414A& 20.71&0.072&$9.612\pm 0.012$&$6.18\pm 0.15$&
$0.741\pm 0.001$&$-1.14\pm 0.07$& 0.45 & 0.36 & SB \\
 94931&G125-04& 28.28&0.030&$8.865\pm 0.005$&$6.13\pm 0.07$&
$0.805\pm 0.005$&$-0.46\pm 0.09$&      &      & IR \\
 95727& 231510& 24.85&0.062&$9.004\pm 0.003$&$5.98\pm 0.13$&
$0.782\pm 0.002$&$-0.44\pm 0.07$& 0.34 & 0.14 &    \\
100568& 193901& 22.88&0.054&$8.652\pm 0.002$&$5.45\pm 0.11$&
$0.555\pm 0.003$&$-1.00\pm 0.07$& 0.35 &      &    \\
112811& 216179& 16.66&0.086&$9.333\pm 0.003$&$5.44\pm 0.18$&
$0.684\pm 0.002$&$-0.66\pm 0.07$& 0.45 & 0.29 &    \\
\\
\multicolumn{11}{c}{Stars in Reid's list}\\
\\
 57450&G176-53& 13.61 &0.113&$ 9.92\pm 0.03$&$5.47\pm 0.25$&
$0.55\pm 0.01$&$-1.26\pm 0.07$&        &      &    \\
 89215&G140-46& 17.00 &0.112&$10.43\pm 0.03$&$6.47\pm 0.24$&
$0.73\pm 0.01$&$-1.51\pm 0.16$&        &      & IR \\
 98020& 188510& 25.32 &0.046&$8.830\pm 0.003$&$5.83\pm 0.10$&
$0.599\pm 0.023$&$-1.37\pm 0.07$& 0.49 &      & SB \\
 99267&G125-64& 12.02 &0.094&$ 9.99\pm 0.06$&$5.31\pm 0.20$&
$0.47\pm 0.02$&$-1.81\pm 0.16$&        &      &E(B-V)=0.04\\
103269&G212-07& 14.24 &0.103&$10.18\pm 0.06$&$5.85\pm 0.22$&
$0.59\pm 0.02$&$-1.48\pm 0.16$&        &      &E(B-V)=0.03\\
106924&G231-52& 15.20 &0.080&$10.19\pm 0.06$&$6.04\pm 0.17$&
$0.58\pm 0.02$&$-1.60\pm 0.16$&        &      &E(B-V)=0.05\\
\\
\multicolumn{11}{c}{Additional stars}\\
\\
   171& 224930& 80.63&0.038&$5.80 \pm 0.01 $&$5.33\pm 0.08$&
$0.665\pm 0.010$&$-0.85\pm 0.07$&      &      & AB \\  
  5336&   6582&132.40&0.005&$5.170\pm 0.031$&$5.79\pm 0.03$&
$0.704\pm 0.004$&$-0.87\pm 0.07$&      &      & SB \\
 17666& 23439A& 40.83&0.055&$8.185\pm 0.013$&$6.24\pm 0.12$&
$0.755\pm 0.009$&$-1.16\pm 0.13$&      &      & SB \\
 18915&  25329& 54.14&0.020&$8.506\pm 0.001$&$7.17\pm 0.04$&
$0.863\pm 0.003$&$-1.69\pm 0.07$&      &      &    \\
 24316&  34328& 14.55&0.069&$9.47 \pm 0.04 $&$5.28\pm 0.15$&
$0.492\pm 0.012$&$-1.44\pm 0.07$&      &      &    \\ 
 39157&  65583& 59.52&0.013&$6.99 \pm 0.01 $&$5.87\pm 0.03$&
$0.762\pm 0.008$&$-0.50\pm 0.13$&      &      & SB \\
 57450&G176-53& 13.61&0.113&$9.92 \pm 0.02 $&$5.59\pm 0.25$&
$0.566\pm 0.016$&$-1.26\pm 0.07$&      &      &    \\
 70681& 126681& 19.16&0.075&$9.31 \pm 0.03 $&$5.72\pm 0.16$&
$0.603\pm 0.003$&$-1.09\pm 0.07$&      &      &    \\
 79537& 145417& 72.75&0.011&$7.531\pm 0.001$&$6.84\pm 0.02$&
$0.815\pm 0.006$&$-1.15\pm 0.13$&      &      &    \\
\tableline
\end{tabular}
\end{table}

\small

\noindent
\begin{table}
\caption{Reddening for the programme globular clusters}
\begin{tabular}{ccccccc}
\tableline
\tableline
NGC&Other&Stars&$E(B-V)_{uvby}$&$E(B-V)_{RHS}$&$E(B-V)_Z$&$<E(B-V)>$\\
\tableline
6341& M92    &   &                 & 0.03 & $0.02\pm 0.01$ & $0.025\pm 0.005$\\
4590& M68    &   &                 & 0.05 & $0.03\pm 0.03$ & $0.040\pm 0.010$\\
7099& M30    & 1 & $0.036\pm 0.024$& 0.04 & $0.04\pm 0.03$ & $0.039\pm 0.001$\\
6205& M13    &   &                 & 0.02 & $0.02\pm 0.02$ & $0.020\pm 0.000$\\
6752&        &   &                 & 0.03 & $0.04\pm 0.02$ & $0.035\pm 0.005$\\
 362&        & 5 & $0.058\pm 0.011$& 0.05 & $0.06\pm 0.03$ & $0.056\pm 0.003$\\
5904& M5     &   &                 & 0.04 & $0.03\pm 0.02$ & $0.035\pm 0.005$\\
 288&        & 7 & $0.026\pm 0.009$&      & $0.04\pm 0.03$ & $0.033\pm 0.007$\\
 104& 47~Tuc & 6 & $0.064\pm 0.010$& 0.06 & $0.04\pm 0.02$ & $0.055\pm 0.007$\\
\tableline
\end{tabular}
\end{table}

\newpage

\noindent
\begin{table}
\caption{Distance moduli for the programme globular clusters}
\begin{tabular}{cccccccccc}
\tableline
\tableline
NGC&Other&[Fe/H]& range  &No.&$(m-M)_V$& cmd   &$(m-M)_V$&No.&$(m-M)_V$\\
   &     &      &        &Stars&       & source&bin cor. &Stars&all sample\\
\tableline
6341& M92&$-2.15$&$-2.5\div -1.5$&2&$14.82\pm 0.08$&1  &14.80& 4&$14.80\pm 0.06$\\
4590& M68&$-1.95$&$-2.5\div -1.5$&2&$15.33\pm 0.08$&2  &15.31& 4&$15.30\pm 0.06$\\
7099& M30&$-1.88$&$-2.5\div -1.3$&3&$14.96\pm 0.08$&3,4&14.94& 7&$14.85\pm 0.04$\\
6205& M13&$-1.41$&$-1.8\div -1.0$&9&$14.47\pm 0.07$&5  &14.45&14&$14.44\pm 0.03$\\
6752&    &$-1.43$&$-1.8\div -1.0$&9&$13.34\pm 0.07$&6  &13.32&14&$13.31\pm 0.03$\\
 362&    &$-1.12$&$-1.6\div -0.8$&6&$15.06\pm 0.08$&7  &15.04&13&$15.03\pm 0.03$\\
5904& M5 &$-1.10$&$-1.6\div -0.8$&7&$14.62\pm 0.07$&8  &14.60&14&$14.62\pm 0.03$\\
 288&    &$-1.05$&$-1.6\div -0.8$&6&$14.96\pm 0.08$&9  &14.94&13&$14.97\pm 0.03$\\
104&47Tuc&$-0.67$&$-1.3\div -0.5$&8&$13.64\pm 0.08$&10 &13.62&13&$13.66\pm 0.03$ \\
\tableline
\end{tabular}
\tablerefs{CMD sources: 1. Stetson \& Harris (1988) 2. McClure et al. (1987)
3. Bolte (1987b) 4. Richer, Fahlman \& VandenBerg (1988) 5. Richer
\& Fahlman (1986) 6. Penny \& Dickens (1986) corrected according to
VandenBerg, Bolte \& Stetson (1990) 7. Bolte (1987a) corrected according to 
VandenBerg, Bolte \& Stetson (1990) 8. Sandquist et al. (1996) 9. Buonanno et
al. (1989) 10. Hesser et al. (1987)} 
\end{table}

\newpage

\noindent
\begin{table}
\caption{Absolute magnitude of the HBs for the programme globular clusters}
\begin{tabular}{cccc}
\tableline
\tableline
NGC&Other&$V(HB)$&$M_V(HB)$ \\
\tableline
6341& M92 &$15.05\pm 0.07$&$0.25\pm 0.10$\\
4590& M68 &$15.71\pm 0.07$&$0.40\pm 0.11$\\
7099& M30 &$15.20\pm 0.10$&$0.26\pm 0.13$\\
6205& M13 &$14.95\pm 0.15$&$0.50\pm 0.17$\\
6752&     &$13.75\pm 0.15$&$0.43\pm 0.17$\\
 362&     &$15.43\pm 0.10$&$0.39\pm 0.13$\\
5904& M5  &$15.11\pm 0.05$&$0.51\pm 0.09$\\
 288&     &$15.40\pm 0.10$&$0.46\pm 0.13$\\
 104&47Tuc&$14.10\pm 0.15$&$0.48\pm 0.17$\\
\tableline
\end{tabular}
\end{table}

\newpage

\noindent
\begin{table}
\caption{Ages for the programme globular clusters}
\begin{tabular}{ccccccccccl}
\tableline
\tableline
NGC&Other&[Fe/H]&$V(TO)$&$M_V(TO)$&SC96&VdB97&DCM97&DCM97&B97&HB \\
   &     &      &       &         &MLT &MLT  &MLT  & CM  &MLT\\ 
   &     &      &       &         &(Gyr)&(Gyr)&(Gyr)&(Gyr)&(Gyr)\\ 
\tableline
6341& M92 &$-2.15$&18.70&3.90& 13.7 & 13.8 & 14.5 & 13.7 & 12.4 & OoII\\
4590& M68 &$-1.95$&19.10&3.79& 11.4 & 11.4 & 11.9 & 11.5 & 10.1 & OoII\\
7099& M30 &$-1.88$&18.73&3.79& 11.1 & 11.1 & 11.6 & 11.3 & ~9.8 & OoII\\
6205& M13 &$-1.41$&18.50&4.05& 12.5 & 12.1 & 13.0 & 12.4 & 11.3 & BHB \\
6752&     &$-1.43$&17.40&4.08& 12.9 & 12.5 & 13.3 & 13.0 & 11.8 & BHB \\
 362&     &$-1.12$&18.85&3.81& ~9.0 & ~8.6 & ~9.3 & ~9.1 & ~7.8 & OoI\\
5904& M5  &$-1.10$&18.60&4.00& 10.8 & 10.3 & 11.0 & 10.6 & ~9.6 & OoI\\
 288&     &$-1.05$&19.00&4.06& 11.3 & 10.7 & 11.3 & 11.1 & 10.1 & BHB \\
 104&47Tuc&$-0.67$&17.75&4.13& 10.8 & 10.1 & 10.5 & 10.4 & ~9.6 & RHB \\
\tableline
\end{tabular}
\end{table}

\newpage

\noindent
\begin{table}
\caption{Mean age for {\it bona fide} old globular clusters}
\begin{tabular}{lccc}
\tableline
\tableline
Isochrone set                       &$M_{V~\odot}$& Age (1)& Age (2) \\
                                    &          & (Gyr)   & (Gyr)     \\
\tableline
\multicolumn{2}{l}{Mixing Length Theory}\\
D'Antona, Caloi \& Mazzitelli (1997)&4.78&$12.6\pm 0.5$&$12.0\pm 0.5$\\
Straniero \& Chieffi (1996)         &4.82&$12.2\pm 0.4$&$12.2\pm 0.4$\\
VandenBerg (1997)                   &4.83&$11.9\pm 0.5$&$12.0\pm 0.5$\\
Bertelli et al. (1997)              &4.85&$10.9\pm 0.4$&$11.3\pm 0.4$\\
\multicolumn{2}{l}{Canuto-Mazzitelli Theory}\\
D'Antona, Caloi \& Mazzitelli (1997)&4.78&$12.2\pm 0.4$&$11.6\pm 0.4$\\
\tableline
\end{tabular}
\end{table}

\newpage

\newpage

\newpage

\newpage

\newpage

\newpage

\newpage

\newpage

\newpage

\newpage

\end{document}